\documentclass[11pt,a4paper]{article}
\usepackage{jheppub}

\usepackage[english]{babel}
\usepackage[utf8]{inputenc}

\usepackage{marginnote}
\usepackage[normalem]{ulem}
\usepackage{epsfig}
\usepackage{tabu}
\usepackage{boldline}
\usepackage{xspace}
\usepackage{slashed}
\usepackage{multirow}
\usepackage{diagbox}
\usepackage[title]{appendix}
\usepackage{subcaption}
\usepackage[table]{xcolor}
\usepackage{enumitem}
\usepackage[utf8]{inputenc}
\usepackage{colortbl}

\hypersetup{colorlinks,bookmarksopen,bookmarksnumbered,
linkcolor=blue,urlcolor=black,citecolor=red}

\newcommand{\fracchange}[1]{\frac{\Delta #1}{#1} }

\graphicspath{{figs/}}

\subheader{\footnotesize
        \vspace*{-3.7em}
        \begin{flushright}
                DESY-19-234 \\
                CERN-TH-2020-091 \\
                INR-TH-2020-001
        \end{flushright}
}

\title{BBN constraints on universally-coupled ultralight scalar dark matter}

\author[a,b,c]{Sergey Sibiryakov,}
\author[d,e]{Philip Sørensen,}
\author[f]{and Tien-Tien Yu}

\emailAdd{sergey.sibiryakov@cern.ch}
\emailAdd{philip.soerensen@desy.de}
\emailAdd{tientien@uoregon.edu}

\affiliation[a]{\small Institute of Physics,  LPTP, Ecole Polytechnique Fédérale de Lausanne (EPFL),\\ CH-1015 Lausanne, Switzerland}
\affiliation[b]{\small Theoretical Physics Department, CERN, Geneva, Switzerland}
\affiliation[c]{\small Institute for Nuclear Research of the Russian Academy of Sciences, \\
60th October Anniversary Prospect, 7a, 117312 Moscow, Russia}
\affiliation[d]{ \small DESY, Notkestra{\ss}e 85, 22607 Hamburg, Germany}
\affiliation[e]{\small II. Institute of Theoretical Physics, Universit\"{a}t  Hamburg, 22761 Hamburg, Germany}
\affiliation[f]{\small Department of Physics and Institute for Fundamental Science, University of Oregon,\\
Eugene OR 97403 USA
}
\date{\today}

\abstract{Ultralight scalar dark matter can interact with all massive Standard Model particles through a universal coupling. Such a coupling modifies the Standard Model particle masses and affects the dynamics of Big Bang Nucleosynthesis. We model the cosmological evolution of the dark matter, taking into account the modifications of the scalar mass by the environment as well as the full dynamics of Big Bang Nucleosynthesis. We find that precision measurements of the helium-4 abundance set stringent constraints on the available parameter space, and that these constraints are strongly affected by both the dark matter environmental mass and the dynamics of the neutron freeze-out. Furthermore, we perform the analysis in both the Einstein and Jordan frames, the latter of which allows us to implement the model into numerical Big Bang Nucleosynthesis codes and analyze additional light elements. The numerical analysis shows that the constraint from helium-4 dominates over deuterium, and that the effect on lithium is insufficient to solve the lithium problem. Comparing to several other probes, we find that Big Bang Nucleosynthesis sets the strongest constraints for the majority of the parameter space.}

\arxivnumber{2006.04820}

\begin{document}
\maketitle

\newpage
\section{Introduction}
The nature of dark matter (DM) is still unknown, despite many decades
of dedicated research. 
Proposed DM candidates span a mass range from 
$\sim 10^{-22}{~\rm eV}$ to the Planck mass for elementary particles
and further up to several solar masses for composite objects, such as
primordial black holes. 
A well-motivated class of DM candidates is found at the lower end of
this mass range. Here, the DM is represented by an ultra-light 
--- of mass well below $1{~\rm eV}$ ---
boson
with extremely weak couplings to the Standard Model (SM) fields.  
Representatives of this class are 
the axion, which was proposed as a
solution to the strong CP problem of
QCD~\cite{Peccei:1977hh,Peccei:1977ur,Weinberg:1977ma,Wilczek:1977pj},
string theory moduli~\cite{Svrcek:2006yi,Arvanitaki:2009fg}
and dark photons~\cite{Nelson:2011sf,Graham:2015rva}.
Such an ultra-light DM
(ULDM) candidate is
characterized by large particle occupation numbers in phase space 
and can be
described as a classical field.  

The weakness of the coupling between ULDM and the SM makes detecting
ULDM a challenging prospect. Many efforts have been undertaken to
identify observations and experiments that are sensitive to ULDM,
ranging from cosmological and astrophysical
probes~\cite{Marsh:2015xka} to terrestrial
experiments~\cite{Battaglieri:2017aum}. 
One promising avenue for probing ULDM is to look for its imprints in
the abundance of the primordial elements. The formation of the
primordial elements, such as helium, deuterium, and
lithium, is governed by Big-Bang Nucleosynthesis (BBN), which
describes an epoch of primordial nuclear transformations in the
expanding Universe. 
Measurements of the primordial abundances of helium-4 ($^4$He) and deuterium (D)
are very
precise and agree well with the theoretical predictions within the SM,
whereas the inferred abundance of lithium is smaller by about a factor of 3
than the theoretical one~\cite{Pitrou:2018cgg,Fields:2019pfx}. The
latter discrepancy may be due to a systematics in the observations or
some kind of new physics. Putting this lithium problem aside,  
even a mild modification to the conditions in the
early Universe would generically 
lead to observable deviations in the abundances
of $^4$He and D making BBN a powerful probe for
beyond the SM physics~\cite{Iocco:2008va,Pospelov:2010hj}.
In particular, it has been extensively used to constrain the 
variations of fundamental
constants~\cite{Kolb:1985sj,Campbell:1994bf,Coc:2006sx,Berengut:2009js,Alvey:2019ctk} 
and modified theories of 
gravity~\cite{Damour:1998ae,Coc:2006rt,Coc:2008yu,Nakamura:2017qtu,Chen:2019kcu}; a
summary of many of these studies can be found in~\cite{Uzan:2010pm}.  

In this work, we investigate the constraints imposed by 
BBN on the interaction of a real scalar ULDM field $\phi$
with the SM. We focus on the case of a {\em universal} coupling,
meaning that $\phi$ interacts with the SM fields through an effective
metric,
\begin{equation} 
\label{effmetr}
\bar g_{\mu\nu}=g_{\mu\nu}\big(1+2\alpha(\phi)\big)\;.
\end{equation}
This
case is interesting since it preserves the weak equivalence principle (WEP)
in the SM sector and thereby avoids many laboratory bounds. This type of coupling naturally appears in scalar-tensor theories of gravity~\cite{Will:2014kxa}. Further,
we will assume that the function $\alpha(\phi)$ is even, so that its
Taylor expansion starts with the {\em quadratic} term,
\begin{equation}
\label{alphaphi}
\alpha(\phi)\simeq \pm \phi^2/\Lambda^2\; , 
\end{equation}
where $\Lambda$ is a scale of the underlying UV-physics. 
At the fundamental level, this
property can be enforced by requiring the symmetry of the theory under
the reflection $\phi\mapsto -\phi$. As a consequence, the linear
term in the coupling is absent and this scenario avoids the bounds coming from the
tests of $r^{-2}$ fall-off of the gravitational
force~\cite{Adelberger:2003zx}
as well as from
the bending of light by the Sun~\cite{Bertotti:2003rm}. 
Consequently, this type of coupling is only weakly constrained by the
present-day observations. The prospects to probe it using binary
pulsar timing have been discussed in~\cite{Blas:2016ddr,Blas:2019hxz}.   

The amplitude of $\phi$ is diluted by the expansion of the
Universe and hence had a much larger value at earlier
epochs. Correspondingly, the coupling $\alpha(\phi)$ was also 
larger at early times than it is now, with a more pronounced effect on the dynamics of the SM sector. This makes BBN a sensitive probe of the above
scenario. 

Setups similar to ours have been previously considered in the
literature. Refs.~\cite{Damour:1998ae,Coc:2006rt} studied the effects
on the BBN in scalar-tensor gravity theories. However, the scalar field in
these works was assumed to be massless and hence could not play the
role of DM. The non-zero mass was included in~\cite{Stadnik:2015kia}
and more recently 
in~\cite{Belokon:2018hrn}. Our work complements these previous studies 
in three key aspects: 1. We
fully take into account the back-reaction from the SM particles on DM, 
which modifies the DM mass and leaves a substantial 
impact on its evolution, 2. We use a kinetic description of neutron freeze-out, instead of  
the instantaneous approximation, which we find to be inadequate, 3. We perform a numerical BBN analysis
using the {\tt AlterBBN} package~\cite{Arbey:2018zfh}, which allows us to calculate the primordial abundances for all the light elements. This third point is
facilitated by considering the theory in the Jordan frame, where the
effects of $\phi$ reduce to a change in the expansion rate of the
Universe. We also clarify the relation between the calculations in the
Jordan and Einstein frames.

The paper is organized as follows: We define the model in 
section~\ref{sec:scalar}, where we describe our assumptions and give a
preliminary discussion of the BBN sensitivity.  
In section~\ref{sec:DMevolution}, we quantify the cosmological
evolution of the scalar field starting with the assumption that $\phi$
makes up the DM today.
The detailed analysis of the BBN constraints is performed in 
section~\ref{sec:effBBN}. We first analytically estimate the effects
of $\phi$ on the $^4$He abundance in 
section~\ref{sec:analytic} and then present our numerical
constraints in section~\ref{sec:Jordan}.
These are compared to other existing bounds on the model 
in section~\ref{sec:addconstraints}.
We conclude in section~\ref{sec:conclusions}.

\section{A real scalar with universal couplings}
\label{sec:scalar}
We begin by adding to the SM a real scalar field $\phi$ with the Lagrangian
\begin{gather}
\mathcal{L}_\phi = \frac{1}{2} (\partial_\mu \phi)^2-\frac{1}{2}m_\phi^2\phi^2\, .
\end{gather}
We assume that any possible self-interaction of $\phi$ can be
neglected, whereas its couplings to the SM fields preserve the
WEP. This implies that $\phi$ is coupled to the SM through the
effective metric (eq.~\ref{effmetr}). We will further assume the coupling
to be weak, $|\alpha(\phi)|\ll 1$ at the values of the field that we
are going to consider. This condition sets the domain of validity of
all the equations derived below. 

As the SM Lagrangian contains fermions, one needs to specify the form
of the effective vielbein, rather than just the effective metric, 
$\bar e_\mu^a=e_\mu^a\big(1+\alpha(\phi)\big)$. In this way we arrive
at the total Lagrangian of our setup,
\begin{align}
\sqrt{-g}\mathcal{L}_{\rm tot} = &\sqrt{-g}\bigg\{
-\frac{M_{\rm pl}^2}{2}R\left[g_{\mu\nu}\right] +
\frac{g^{\mu\nu}}{2}\partial_\mu\phi \partial_\nu\phi 
-\frac{1}{2}m_\phi^2\phi^2\bigg\} \notag \\
&+\det\big[e_\mu^a\big(1+\alpha(\phi)\big)\big] 
\mathcal{L}_{\rm SM} \left[ e_\mu^a\big(1+\alpha(\phi)\big),\psi\right]\, 
,\label{EinsteinLag}
\end{align}
where $M_{\rm pl}$ is the Planck mass, $R$ is the Ricci scalar, and $\mathcal{L}_{\rm SM}$ is the SM Lagrangian and we have collectively
denoted the SM fields with $\psi$. Note that we have assumed the
standard Einstein--Hilbert term for the gravitational action.
Expanding to the linear order in $\alpha$ we can write the leading
interaction term as
\begin{equation}
\label{Lint}
\mathcal{L}_{\rm int}=-\alpha(\phi) \Theta_{\rm SM}[\psi]\;,
\end{equation}
where $\Theta_{\rm SM}[\psi]$ is the trace of the energy-momentum
tensor (EMT) of the SM sector.

Let us discuss the contributions of various fields to
eq.~\ref{Lint}. For fundamental fermions (quarks and leptons), using
their equations of motion, we obtain,
\begin{equation}
\label{Lintf}
\mathcal{L}^f_{\rm int}=-\alpha(\phi) m_f\bar f\,f\;.
\end{equation}
Similarly, for massive vector fields (W and Z bosons) we have,
\begin{equation}
\label{LintV}
\mathcal{L}^V_{\rm int}=\alpha(\phi) m_V^2 V_\mu\,V^\mu\;.
\end{equation}
In both cases the interaction amounts to a rescaling of the particle
mass by a $\phi$-dependent factor,
\begin{equation}
\label{massresc}
m_{f,V}\mapsto m_{f,V}\big(1+\alpha(\phi)\big) \;.
\end{equation}
The situation is more subtle for composite particles, such as hadrons,
whose mass is dominated by non-perturbative QCD
contributions. Nevertheless, we presently argue that the mass scaling
(eq.~\ref{massresc}) applies to them as well. Indeed, the non-perturbative
contributions to the hadron masses are determined by the QCD scale
$\Lambda_{\rm QCD}$. The latter can be related to the physics at high
energies using the renormalization group running of the strong
coupling constant $\alpha_s$. Including the mass thresholds from
charm, bottom and top quarks, one has~\cite{Coc:2006sx},
\begin{equation}  
\label{LQCD}
\Lambda_{\rm QCD}=M_{\rm UV}\bigg(\frac{m_c m_b m_t}{M_{\rm
    UV}^3}\bigg)^{2/27}
\exp\bigg(-\frac{2\pi}{9\alpha_s(M_{\rm UV})}\bigg)\;,
\end{equation}
where $m_{c,b,t}$ are the masses of the charm, bottom, and top quarks, respectively, and $M_{\rm UV}>m_t$ is the high-energy scale where the strong coupling
is normalized to a given value. If the $\phi$ interactions are to
preserve the WEP, the scale $M_{\rm UV}$ must vary according to 
\begin{equation}
\label{MUVvar}
M_{\rm UV}\mapsto M_{\rm UV}\big(1+\alpha(\phi)\big)\;,
\end{equation}
 with $\alpha_s\big(M_{\rm UV}(1+\alpha)\big)$ held fixed. This will be
the case, for example, in Grand Unified Theories where the cutoff scale
$M_{\rm UV}$ corresponds to a physical particle mass. Then
$\Lambda_{\rm QCD}$ scales in the same way as the fundamental particle
masses, 
\begin{equation}
\label{LQCDvar}
\Lambda_{\rm QCD}\mapsto \Lambda_{\rm QCD}\big(1+\alpha(\phi)\big)\;,
\end{equation}
implying that the $\phi$-dependence of all hadronic quantities
is dictated by their mass dimensions. This applies not only to the
hadron masses, but also to nuclear binding energies, decay widths, and
cross sections. See appendix~\ref{app:variations} for more discussion of this point. 

In principle, one could envision scenarios where the UV scale does not
obey eq.~\ref{MUVvar}. However, this will inevitably lead to a violation
of WEP in the low-energy physics and would make the predictions of the
theory dependent on the details of the UV completion. In this paper we
avoid such complications by restricting to the WEP preserving case.

We can get another perspective on the effects due to $\phi$ by
transforming the metric from the Einstein to Jordan frame,
$g_{\mu\nu}\mapsto\bar g_{\mu\nu}$. Keeping the leading terms in
$\alpha$ we obtain from eq.~\ref{EinsteinLag},
\begin{align}
\sqrt{-g}\mathcal{L}_{\rm tot}=\sqrt{-\bar g}\bigg\{&
-\frac{M^2_{\rm pl}}{2}\big(1-2\alpha(\phi)\big)R[\bar g_{\mu\nu}]
 -3 M^2_{\rm pl}\, \bar g^{\mu\nu}\partial_\mu\alpha\partial_\nu
 \alpha\notag \\
&+\frac{\bar g^{\mu\nu}}{2} (1-2\alpha) \partial_\mu\phi\partial_\nu \phi
 -\frac{1}{2}(1-4\alpha) m_\phi^2\phi^2
 +\mathcal{L}_{\rm SM}[\bar e^a_\mu,\psi]\bigg\}\;.
\label{JordanLag1}
\end{align}
Note that the second term in the first line, though quadratic in
$\partial_\mu\alpha$, is enhanced by the Planck mass and cannot be
neglected in general. We observe that $\phi$ has decoupled from the
SM, which now interacts covariantly only with the metric $\bar
g_{\mu\nu}$. Quantum corrections within the matter sector do not
change this covariant form and thus do not spoil WEP. The latter can
still be violated by quantum corrections involving metric
perturbations in the loops. These are, however, suppressed by the
inverse Planck mass.

The non-minimal coupling of $\phi$ to the Jordan metric $\bar g_{\mu\nu}$
implies that we are dealing with a variant of scalar-tensor
gravity. The field $\phi$ affects BBN by modifying the expansion rate
of the universe at the BBN epoch. The modification stems from the
$\phi$-dependence of the Planck mass,
\begin{equation}
\label{Mplvar}
M_{\rm pl}^2\mapsto M_{\rm pl}^2\big(1-2\alpha(\phi)\big)\;,
\end{equation}   
as well as the EMT of $\phi$. Focusing for simplicity on the first effect --- the
variation of the Planck mass
--- we can get a rough idea of the BBN
reach in constraining the model parameters. The difference between the
Planck mass at the time of BBN and today is bounded at the level of $3\%$
\cite{Alvey:2019ctk}. The coupling $\alpha(\phi)$ nowadays is very
small due to the dilution of the amplitude of $\phi$ by the expansion of the universe.
Thus, we expect BBN to exclude the parameter
region where the coupling $|\alpha(\phi)|$ exceeds roughly $0.015$ at
the BBN epoch. On the other hand, BBN will remain insensitive to the
presence of $\phi$ if $|\alpha(\phi)|\ll 0.01$ at $t_{\rm BBN}$. 
We will see below that the presence of the $\phi$ EMT, as well as its time evolution, make the true story somewhat more complicated. In particular, the constraining power can be reduced in some parameter regions due to cancellation between several competing effects.

We now further specify the model by imposing the condition that the coupling
$\alpha$ is an even function of $\phi$, so that its Taylor expansion
starts with a quadratic term (see eq.~\ref{alphaphi}). The higher-order terms are assumed to be negligible as long as
$\phi^2/\Lambda^2\ll 1$. This implies the validity of the 
form
(eq.~\ref{alphaphi}) all the way back through BBN for the relevant range
of parameters. This has an important implication for the dynamics of
$\phi$ in the early Universe. From the expression (eq.~\ref{Lint}) for
the interaction Lagrangian in the Einstein frame we see that the
presence of SM matter with energy density $\rho_{\rm SM}$ and pressure
$p_{\rm SM}$ induces a contribution into the mass term of $\phi$
leading to an effective time-dependent mass,
\begin{equation}
\label{effmass}
m_{\phi,{\rm eff}}^2 = m_\phi^2 \pm 2\frac{\Theta_{\rm
    SM}}{\Lambda^2}~,~~~~~
\Theta_{\rm SM}= \rho_{\rm SM}-3p_{\rm SM}\; .
\end{equation}
In the next section we study the evolution of the ULDM field taking
this effective mass into account.

Prior to concluding this section, let us comment on the issue of quantum
corrections to the $\phi$-mass and its self-interaction due to its
coupling to the SM. Working in the Einstein frame and assuming, as
before, the preservation of WEP we obtain that the result of
integrating out the SM fields should
have the form,
\begin{equation}
\label{deltaLphi}
\sqrt{-g}\,\delta\mathcal{L}_\phi=\det\big[e_\mu^a\big(1+\alpha(\phi)\big)\big]\,
\rho_{\rm SM}^{\rm vac}\;,
\end{equation}
where $\rho_{\rm SM}^{\rm vac}$ is the SM vacuum energy density. It is
well-known that a naive estimate of loop contributions would yield 
$\rho_{\rm SM}^{\rm vac}\sim m_t^4$. Such a large contribution would completely destroy our
scenario. However, we also know that the total vacuum energy density 
$\rho_{\rm tot}^{\rm vac}$ is very small, implying a delicate
cancellation between different contributions into it, including the
bare (unrenormalized) value. The mechanism ensuring this cancellation
is still unknown, which constitutes the famous cosmological constant
problem. We do not attempt to add anything to its solution and just
speculate that the cancellation of vacuum energy can happen separately
within the SM sector (or its extension universally coupled to
$\phi$) implying  $\rho_{\rm SM}^{\rm vac}\lesssim \rho_{\rm tot}^{\rm
  vac}$. In this case the effect of quantum corrections
(eq.~\ref{deltaLphi}) is completely negligible.

\section{Evolution of dark matter}
\label{sec:DMevolution}
To understand the effect of DM on BBN, we calculate the cosmological
evolution of the DM field $\phi$ from the present epoch back to the time 
when the temperature of the Universe was $T\sim {\rm MeV}$. This epoch
corresponds to the freeze-out of the weak interactions and sets the
neutron abundance, which is one of the key ingredients of BBN. The
energy density of $\phi$ is subdominant until the epoch of matter
domination. In addition, we will neglect in this section the $\phi$-induced changes in
the evolution of the SM fields. These changes would lead only to $O(\alpha^2)$ corrections in
the eventual BBN analysis, whereas the leading effects are linear in
$\alpha$. Thus, working in the Einstein frame, we can treat $\phi$ as a probe field embedded into a
Friedmann--Robertson--Walker (FRW) universe with the scale factor
$a(t)$ obeying the standard expansion history. Assuming that the field
is spatially homogeneous, we obtain its equation of motion,
\begin{gather}
\ddot{\phi} + 3H\dot{\phi}+m_{\phi,{\rm eff}}^2\,\phi = 0\, ,\label{Klein-Gordon} 
\end{gather}
where $\dot{\ }$ denotes the derivative with respect to time, $H\equiv \dot a/a$ is the Hubble rate, and $m^2_{\phi,{\rm eff}}$ is given by eq.~(\ref{effmass}). 

\begin{figure}
\centering
\includegraphics[width=0.48\textwidth]{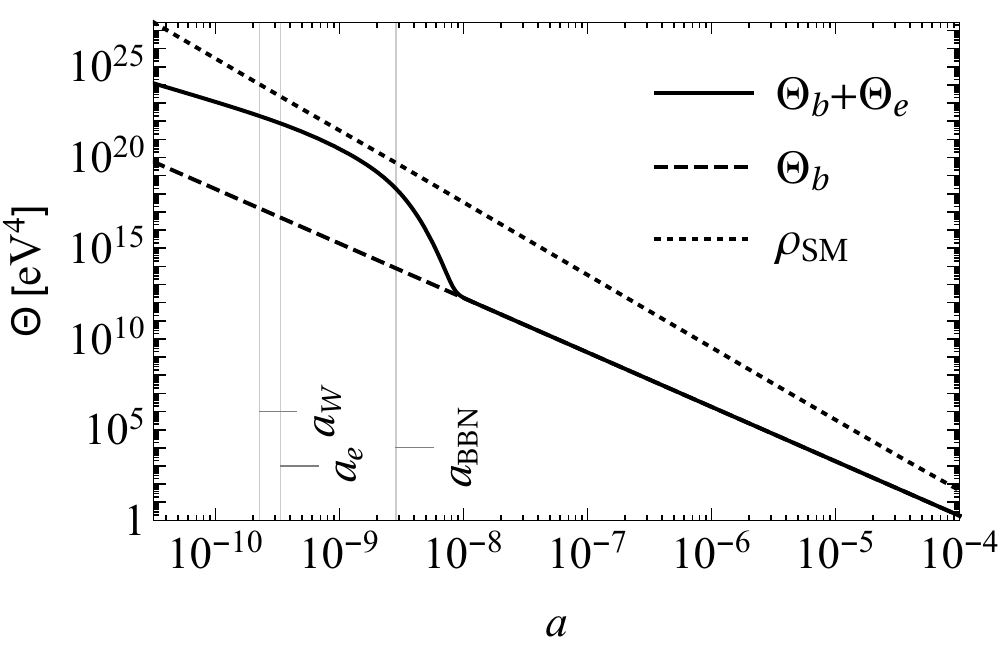}\qquad
\includegraphics[width=0.46\textwidth]{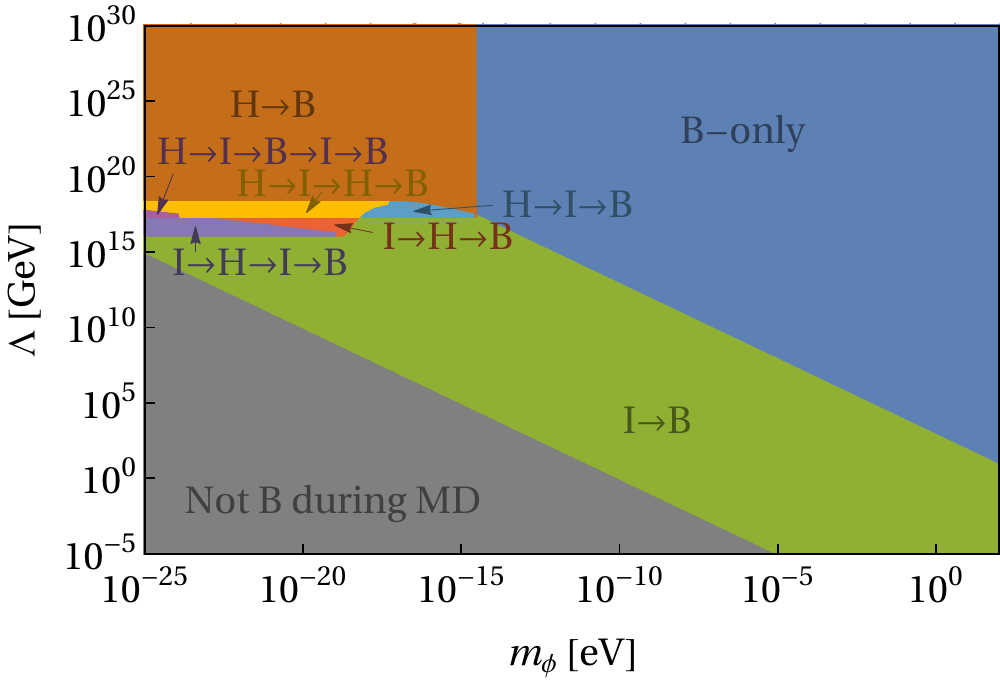}
\caption{\small \textbf{Left:} The evolution of $ \Theta_{\rm SM} $ as a
  function of scale factor (solid line). The contribution of
  non-relativistic baryons $ \Theta_b \propto a^{-3} $
  is displayed by the dashed line for reference. Notice the large contribution from the
  $e^+e^-$ plasma. For reference, we also show the total energy density $\rho_{\rm SM}$ (dotted line). Vertical lines mark the values of the scale factor
 corresponding
  to freeze-out of weak interactions ($a_{\rm W}$), the time when electrons become non-relativistic ($a_{e}$)
  and BBN ($a_{\rm BBN}$).  
\textbf{Right:}
Map of the transition history of DM evolution for various
points in parameter space. Each region is labeled with the regimes the
field $\phi$ passes from the weak freeze-out to the present time. 
Example: $I\to H\to B$ refers to an
evolution that starts out dominated by the induced mass, then 
transitions to being dominated by Hubble friction and
finally by the bare mass, $m_\phi$. The gray-shaded region on the bottom left is excluded by the condition eq.~\ref{constr1}.
} 
\label{EffectOfEPAnnihilation}
\end{figure}

Let us discuss the trace of the SM EMT, which enters into the effective
mass. Relativistic particle species (photons, neutrinos) do not
contribute into it\footnote{We neglect the trace anomaly which is relevant only at temperatures above several
  MeV when electrons are relativistic.}. For most of the time the SM EMT is dominated by the non-relativistic baryonic matter and reads,
\begin{equation}
\label{ThetaB}
\Theta_b(t)=\rho_b(t)=\frac{3 M_{\rm pl}^2H_0^2\Omega_b}{a^3(t)}\;,
\end{equation}  
where $\rho_b(t)$ is the time dependent energy density of baryons,
$H_0$ and $\Omega_b$ are the present-day Hubble constant and the
baryon density fraction, respectively, and we have normalized the scale
factor to be unity today, $a_0=1$. However, around the epoch of
BBN, $\Theta_{\rm SM}$ receives a large additional contribution from the
electron-positron plasma. Due to the numerical coincidence between the
electron mass and the BBN temperature, electrons and positrons become
non-relativistic during BBN, while their number density still greatly
exceeds that of protons and neutrons until $e^+$
annihilate with $e^-$ at somewhat lower temperatures. Using the
standard thermodynamic expressions for the energy density and pressure
of a Fermi gas we obtain the trace of $e^+e^-$ EMT,
\begin{equation}
\label{ThetaE}
\Theta_{e}=4\cdot\frac{m_e^2 T^2}{2\pi^2}
\int_{m_e/T}^{\infty}dx \frac{\sqrt{x^2-m_e^2/T^2}}{e^x+1}\, ,
\end{equation}
where $m_e$ is the electron mass and we have neglected the electron chemical
potential. The factor 4 in front counts the number of spin degrees of
freedom. The total SM EMT trace is given by
\begin{equation}
\label{Thetaall}
\Theta_{\rm SM}=\Theta_b+\Theta_e
\end{equation}
and is shown in fig.~\ref{EffectOfEPAnnihilation} (left panel) as a
function of the scale factor\footnote{To convert the temperature into
  the scale factor we used the formula~\cite{Riotto:2002yw} 
$T=\left(\frac{g_{*S}(T_0)}{g_{*S}(T)}\right)^{1/3}\frac{T_0}{a}$, where $T_0$
is the present-day Cosmic Microwave Background temperature and
$g_{*S}(T)$ is the effective number of relativistic degrees of freedom
in the entropy density, which we obtain from the {\tt microOMEGAs}
package~\cite{Belanger:2013oya}.}.  Note that the EMT trace never exceeds $10\%$ of the SM energy density during the BBN epoch~\cite{Erickcek:2013dea},
\begin{equation}
\label{eq:Sigma}
 \Sigma\equiv \Theta_{\rm SM}/\rho_{\rm SM} <0.1 \;. 
\end{equation}

The evolution of $\phi$ has three distinct regimes 
characterized by the dominant term in the equation of motion eq.~\ref{Klein-Gordon}:
\begin{itemize}
	\item \textbf{Hubble friction dominance} (H): $H^2\gg
          m^2_{\phi,{\rm eff}}$\;,
	\item \textbf{Constant ``bare'' mass dominance} (B): 
$ m_\phi^2\gg H^2,~\Theta_{\rm SM}/\Lambda^2$\;,
	\item \textbf{Induced mass dominance} (I): 
$ \Theta_{\rm SM}/\Lambda^2\gg H^2, m_\phi^2$\;.
\end{itemize}
Depending on the model parameters, the field goes through these
regimes in various sequences, which may be quite complicated, 
as shown in fig.~\ref{EffectOfEPAnnihilation} (right panel). This is due to
the non-trivial time-dependence of the induced mass. 

In the regime (H), the field is frozen at a constant value. On the
other hand, in the (B) regime the field oscillates,
\begin{equation}
\label{Bosc}
\phi=\Phi(t) \cos(m_\phi t+\varphi)\;,
\end{equation}
where $\varphi$ is a constant phase and the amplitude decreases
with time as $\Phi(t)\propto a^{-3/2}(t)$. In this regime the field
$\phi$ behaves as DM~\cite{Marsh:2015xka}. In particular, its 
energy density scales inversely proportional to $a^3$,
\begin{align}
\rho_\phi=\frac{m^2_\phi\Phi^2}{2}\propto \frac{1}{a^3}\;.
\end{align}
Equating this to the measured average DM density today 
$\rho_{{\rm DM},0}=1.26\times 10^{-6}\,{\rm GeV/cm}^3$ 
determines the present-day amplitude 
\begin{equation}
\label{phiampnow}
\Phi_0=\frac{\sqrt{2\rho_{{\rm DM},0}}}{m_\phi}\;.
\end{equation}
To reproduce the success of the $\Lambda$CDM cosmology, the field must
be in the regime (B) throughout the matter-dominated stage of the 
history of the universe, which puts the constraint on the model parameters,
\begin{equation}
\label{constr1}
m_\phi^2\Lambda^2\gg \Theta_{\rm SM}(a_{\rm eq})=
\frac{3 M_{\rm pl}^2H_0^2\Omega_b}{a_{\rm eq}^3}\simeq 1.8\,{\rm eV}^4\;,
\end{equation}
where $a_{\rm eq}\simeq 10^{-4}$ is the scale factor at the epoch of
matter-radiation equality.

The behavior of the field $\phi$ in the regime (I) differs
qualitatively, depending on the sign of the coupling
(\ref{alphaphi}). 

\begin{figure}
\centering
\includegraphics[width=0.48\textwidth]{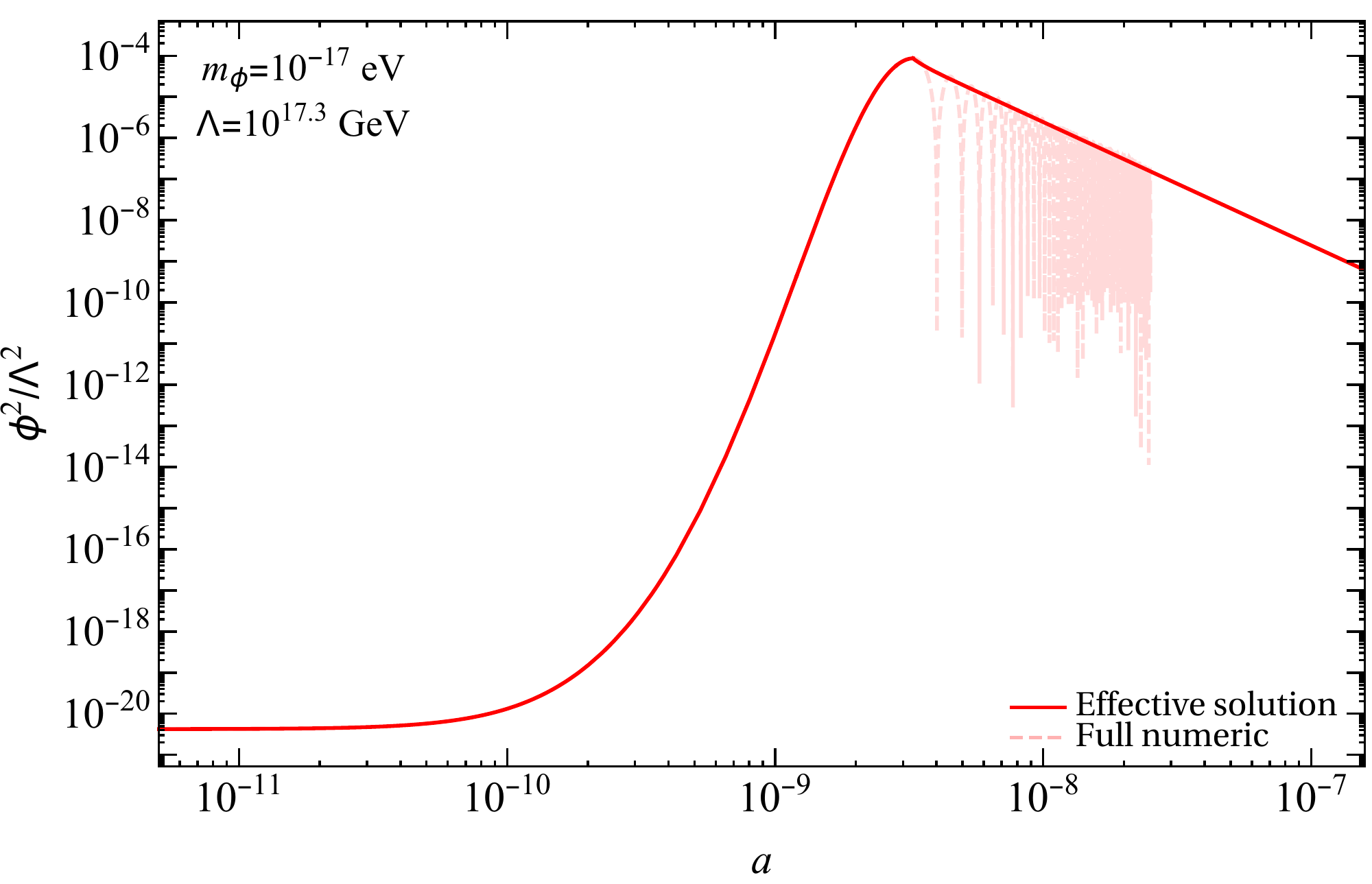}
\quad
\includegraphics[width=0.48\textwidth]{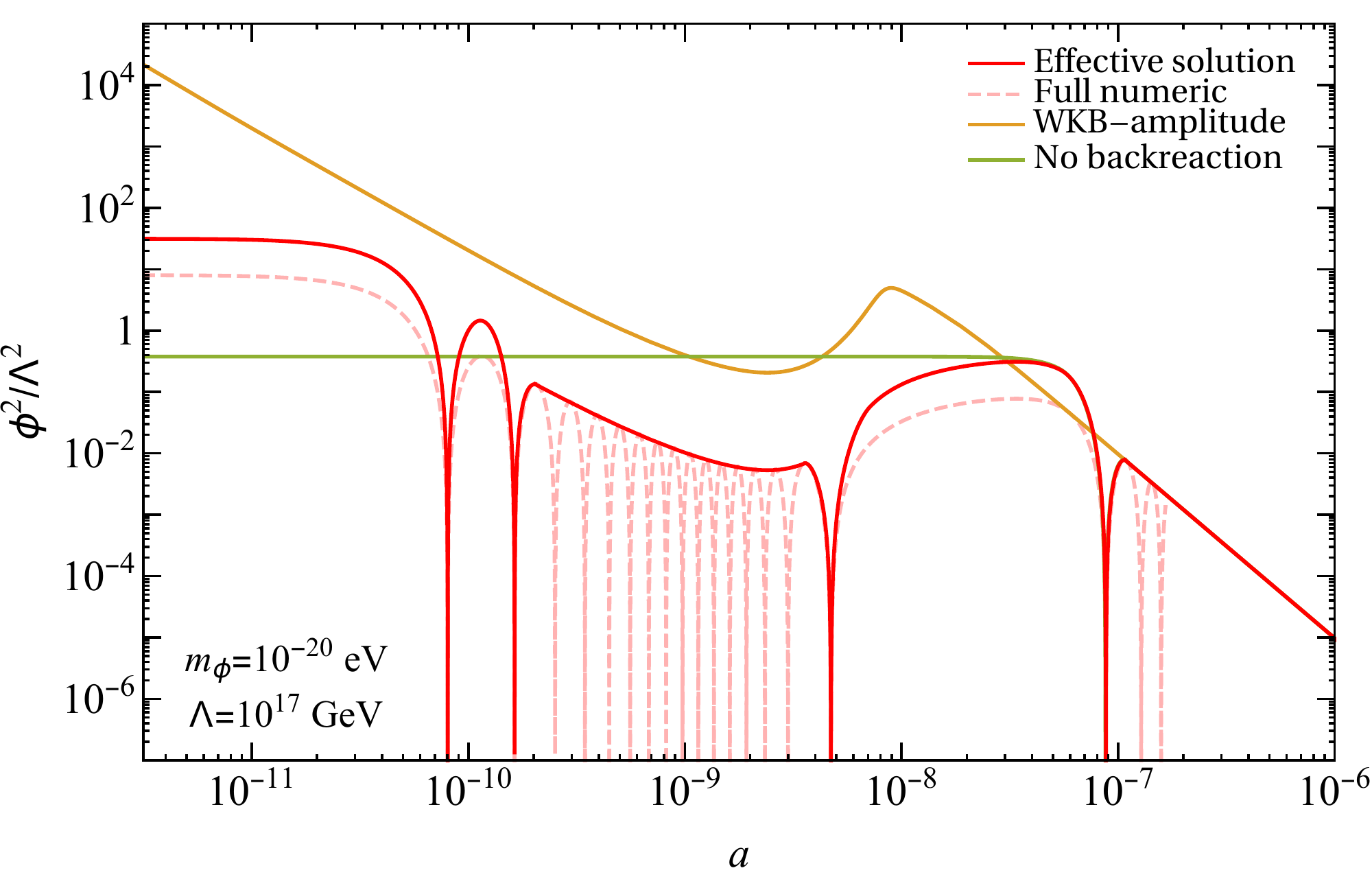}
\caption{\small {\bf Left:} Evolution of $\phi^2/\Lambda^2$ as a function of scale factor $a$ for the negative coupling and parameters $m_\phi=10^{-17}$ eV, $\Lambda=10^{17.3}$ GeV. The red-dashed curve shows the full numeric solution whereas the solid red curve shows the effective solution which patches together the oscillations with an exponential growth.  {\bf Right:} Evolution of $ \phi^2/\Lambda^2$ as a function of scale factor $a$ for positive coupling and parameters $m_\phi=10^{-20}$ eV, $\Lambda=10^{17}$ GeV. The red-dashed curve shows the full numeric solution whereas the red solid curve gives the effective solution which patches together the slowly oscillating/frozen 
phase with a WKB-type solution in the intermediate regime. The pure WKB amplitude, neglecting Hubble friction, is shown in orange. 
The green curve shows the evolution of the $\phi$-field neglecting the induced mass. \label{fig:DM density}}
\end{figure}

\paragraph{Negative coupling} In this case the induced mass term is
tachyonic and leads to an exponential growth of the field, with the
approximate solution,
\begin{equation}
\label{expgrowth}
\phi\propto \frac{1}{a^{3/2}\Theta_{\rm SM}^{1/4}}\,
\exp\bigg(\int dt\,\frac{\sqrt{2\Theta_{\rm SM}}}{\Lambda}\bigg)\;.
\end{equation} 
In principle, this can serve as a mechanism for DM production. Note,
however, that it requires a fine-tuning of the initial $\phi$-value 
in order not to over-produce DM. We do not discuss a possible origin
of such tuning. Instead, we adopt a phenomenological approach and, in
order to see what constraints BBN imposes on this type of coupling, 
evolve the field backward in time from today by matching the
oscillations in the (B) regime to the exponential growth
(\ref{expgrowth}) in the (I) regime. Fig.~\ref{fig:DM density} (left
panel) 
shows
an example of the $\phi$ evolution for $m_\phi=10^{-17}$ eV and $\Lambda=10^{17.3}$ GeV. The red-dashed curve denotes the full numeric solution while the red-solid curve shows the effective solution which patches the oscillations in the (B) regime to the exponential growth in the (I) regime. The details of the patching procedure are given in appendix~\ref{app:DMevolution}.

\paragraph{Positive coupling} 
In this case the solution in the regime (I) is oscillating, similar to
the regime (B). As the oscillations are much faster than the
expansion of the Universe, we can obtain a unified description in
these two regimes using the WKB-type expansion. This yields,
\begin{equation}
\label{WKB}
\phi=\Phi(t)\cos\bigg(\int dt\,m_{\phi,{\rm eff}}(t) +\varphi\bigg)\;,
\end{equation} 
where the amplitude scales as,
\begin{equation}
\label{WKBampl}
\Phi(t)\propto a^{-3/2}(t)\, m_{\phi,{\rm eff}}^{-1/2}(t)\;.
\end{equation} 
Using this scaling and the known amplitude $\Phi_0$ today, we
determine the amplitude back in time until the field enters into the
(H) regime, where it freezes at a constant. To patch the solutions in
(H) and (I)/(B) regimes, we solve eq.~(\ref{Klein-Gordon})
numerically. As already noted, the field $\phi$ may pass between
different regimes several times. In that case, the frozen and oscillatory behavior alternate.
In our BBN analysis, we use an effective solution where $\phi^2/\Lambda^2$ is averaged over fast oscillations while slow oscillations and frozen regions are resolved without averaging. This is obtained by patching the numerical solution with the WKB amplitude for the rapidly oscillating regions. We also take into account the factor $1/2$ which appears due to averaging (see appendix~\ref{app:DMevolution} for details). An illustrative example of the 
$\phi$-evolution for $m_\phi=10^{-20}$ eV and $\Lambda=10^{17}$ GeV is shown in fig.~\ref{fig:DM density} (right panel). 

\section{Effect on BBN}
\label{sec:effBBN}
\subsection{Analytic estimates}
\label{sec:analytic}

To understand the modification of BBN predictions due to ULDM, we
first consider a simplified picture of BBN that will allow us to capture the main physical effects analytically. The full implementation of ULDM
effects in the BBN code {\tt AlterBBN} \cite{Arbey:2018zfh} will be discussed in the next subsection.

\subsubsection{Standard picture}
\label{standard}
We begin with a recap of the key stages of the standard BBN dynamics following~\cite{Mukhanov:2003xs}.  
At early times, neutrons and protons are in thermal equilibrium
through the reactions
$n+\nu_e\leftrightarrow p+e^-$ and $n+e^+\leftrightarrow p+\bar\nu_e$
which results in a neutron abundance (i.e. the ratio of the amount of
neutrons to the total amount of baryons),
\begin{equation}
\label{npweak}
\frac{n_n}{n_b}=\frac{1}{1+e^{m_{np}/T}}\equiv X_n^{\rm eq}\, ,
\end{equation}
where $T$ is the temperature and $m_{np}\simeq 1.29\;{\rm MeV}$
is the difference
between the neutron and proton masses. 
At $T_W\simeq 0.8$~MeV the rates of the weak reactions become
comparable to the Hubble expansion rate and further evolution of the
neutron abundance is governed by the kinetic equation,
\begin{equation}
\label{kinetic}
\frac{dX_n}{dt}=-\lambda_{n\to p}\; \big(1+e^{-m_{np}/T}\big)(X_n-X_n^{\rm eq})\;.
\end{equation}
Here $\lambda_{n\to p}$ is the rate of neutron-to-proton conversion,
which can be approximated as
\begin{equation}
\label{nprate}
\lambda_{n\to p}=\frac{1+3g_A^2}{\pi^3}\; G_F^2\, T^5 J\big(m_{np}/T\big)\;,
\end{equation}
where $G_F$ is the Fermi constant, $g_A\simeq 1.26$ is the nucleon axial
charge, and the phase space
integral has the form,\footnote{For simplicity, 
we assume that neutrinos and electrons
have the same temperature $T>m_e$.}
\begin{equation}
J(z)\simeq \frac{45\zeta(5)}{2}+\frac{7\pi^4}{60}z
+\frac{3\zeta(3)}{2}\bigg(1-\frac{m_e^2}{2m_{np}^2}\bigg)z^2\;.
\end{equation}
By solving the linear equation (\ref{kinetic}) we find the neutron
abundance after the weak freeze-out,
\begin{equation}
\label{Xnweak}
X_{n,{\rm W}}=-\int_0^\infty da \frac{d X_n^{\rm eq}}{da}
\exp\bigg[-\int_a^\infty \frac{da_1}{a_1} 
\frac{\lambda_{n\to p}}{H(a_1)} \Big(1+e^{-\frac{m_{np}}{T}}\Big)\bigg]\;,
\end{equation}
where we have switched from time to the scale factor as the
integration variable for later convenience and the upper limit of integration refers to arbitrarily late times after weak freeze-out. 
The equilibrium neutron abundance $X_n^{\rm eq}$ and the rate
$\lambda_{n\to p}$ in this formula are understood as functions of the
scale factor. 
Evaluating the integral using the standard thermal history yields,
\begin{gather}
X_{n,{\rm W}}\simeq 0.157\,.  \label{weakr}
\end{gather}

After freeze-out, the number of neutrons continues to slowly decrease due to neutron
decay, 
\begin{equation}
\label{npBBN}
X_{n,{\rm BBN}}=X_{n,{\rm W}}
e^{-\Gamma_n(t_{\rm BBN}-t_{\rm W})}\simeq 0.12\;,
\end{equation}
where the neutron lifetime is $\Gamma_n^{-1}\simeq 880\;{\rm s}$ and $t_{\rm W} \sim 1\;{\rm s}$.
The onset of BBN is held back by the efficient dissociation of
deuterium, which is often referred to as the deuterium
bottleneck. This delays BBN until the universe cools to a temperature
which allows deuterium to survive long enough to form heavier
elements. 
This temperature can be estimated from the Saha equation for the
equilibrium abundance of deuterium,
\begin{gather}
X_D\equiv\frac{n_D}{n_b}\simeq 16.3\,\eta_b \,X_pX_n\left( \frac{T}{m_p}
\right)^{3/2} e^{B_D/T} , \label{Saha}
\end{gather}
where $X_p=1-X_n$ is the proton abundance, 
$B_D\simeq 2.22\;{\rm MeV}$ is the deuterium binding energy and 
$\eta_b=(6.104\pm 0.058)\cdot 10^{-10}$ is the baryon-to-photon ratio,
as inferred from the Planck data~\cite{Fields:2019pfx}.
The bottleneck opens when the rate of conversion of deuterium into
heavier elements becomes comparable to the expansion of the
universe. This occurs when its abundance reaches
$X_D\sim 2\times 10^{-5}$~\cite{Mukhanov:2003xs}. At this moment, the deuterium burning is
still much slower than the reaction $n+p\leftrightarrow D+\gamma$
maintaining deuterium in equilibrium with neutrons and protons, and
the deuterium abundance continues to grow until it reaches its maximal
value $X_D^{\rm max}\sim 10^{-2}$. According to eq.~(\ref{Saha}), this
happens at temperature $T_{\rm BBN}\simeq 75$~keV.
Note that it is only logarithmically sensitive to the
the precise value of $X_D^{\rm max}$ and corresponds to the time $t_{\rm BBN}\simeq 230\,{\rm s}$ used in
eq.~(\ref{npBBN}).  
At this epoch the deuterium burning is very fast and its abundance
drops, whereas the majority of available neutrons end up
in $^4$He due to its high binding energy, resulting in the primordial 
$^4$He mass fraction of
\begin{gather}
Y_{\rm p} = 2 X_{n,\rm BBN} \simeq 0.24\, .\label{standard Yp}
\end{gather}

This estimate agrees well with the precise calculation using numerical 
codes \cite{Fields:2019pfx},
\begin{subequations}
\begin{equation}
Y_{\rm p}^{\rm th}=0.24691\pm 0.00018\;. 
\end{equation}
The standard theory also predicts the primordial abundances of other
light elements, such as deuterium and lithium
\cite{Fields:2019pfx},\footnote{We do not discuss $^3$He as the
  experimental determination of its abundance is currently subject to
  large uncertainties~\cite{Tanabashi:2018oca}.}
\begin{align}
({\rm D}/{\rm H})_{\rm p}^{\rm th}=(2.57\pm 0.13)\cdot 10^{-5}\;,\qquad\qquad
({\rm Li}/{\rm H})_{\rm p}^{\rm th}=(4.72\pm 0.72)\cdot 10^{-10}\;,
\end{align}
\end{subequations}
The experimentally measured values of these abundances inferred from
astrophysical observations are~\cite{Tanabashi:2018oca},
\begin{subequations}
\begin{align}
&Y_{\rm p}^{\rm exp}=0.245\pm 0.003\;,\\
&({\rm D}/{\rm H})_{\rm p}^{\rm exp}=(2.547\pm 0.025)\cdot 10^{-5}\;,\\
&({\rm Li}/{\rm H})_{\rm p}^{\rm exp}=(1.6\pm 0.3)\cdot 10^{-10}\;.
\end{align}
\end{subequations}
We see that for $^4$He and D the theory and experiment are in
very good agreement. In particular, for $^4$He the relative difference
is 
\begin{gather}
\frac{\Delta Y_{\rm p}}{Y_{\rm p}} = \frac{Y_{\rm p}^{\rm exp}-Y_{\rm
    p}^{\rm th}}{Y_{\rm p}^{\rm th}} = -0.008\pm 0.012 \, , 
\label{He4 constraint}
\end{gather}
where the error is dominated by the experimental uncertainty. 
This relative difference constrains how much the addition of dark
matter is allowed to change the SM prediction. On the other hand, for
Li the theoretical prediction is almost 3 times higher than 
the observed value --- the mismatch is known as the ``lithium
problem''. For constraints on ULDM, we 
will use only $^4$He
and D abundances.

\subsubsection{Effects of dark matter on Helium-4 abundance}
\label{sec:He4analytical}

The abundance of $^4$He is of particular interest as it depends only
on a few factors. We now derive an analytic estimate of the ULDM
impact on it using the formulae from the previous subsection. To this
end, we consider the relative change
\begin{gather}
\frac{\Delta Y_{\rm p}}{Y_{\rm p}} \simeq \frac{\Delta X_{n,{\rm
      W}}}{X_{n,{\rm W}}} -
\Delta\left(\int_{a_{\rm W}}^{a_{\rm BBN}}\frac{da}{aH(a)}\Gamma_n(a)\right)\, ,
\label{He4 fraction}
\end{gather}
where we have used eqs.~(\ref{npBBN}), (\ref{standard Yp}) and have written the
factor describing neutron decay as an integral over the scale factor
between the weak freeze-out and BBN. Note that we have allowed the
neutron decay width to be time-dependent due to its modulation by ULDM.
We first work in the Einstein frame and then present the 
Jordan frame description.

\paragraph{Einstein frame} As discussed in sec.~\ref{sec:scalar} (see also
appendix~\ref{app:variations}), the universal coupling of DM to SM affects the
masses of particles and other dimensionful quantities, while leaving 
the dimensionless ratios intact. The Fermi constant has mass dimension
$-2$, whereas 
the neutron-proton mass difference, the neutron width and the deuterium
binding energy all have unit mass dimension. Thus, we have, 
\begin{equation}
\label{masschanges}
-\frac{1}{2}\fracchange{G_F}=\fracchange{m_{np}}=\fracchange{\Gamma_n}=
\fracchange{B_D}=\alpha(\phi)\, .
\end{equation}

Next, it is shown in appendix~\ref{app:Einstein-Jordan map} that the SM
energy density and the Hubble rate in the Einstein frame remain the
same functions of the scale factor, as in the standard cosmology,
\begin{equation}
\label{rhoHEins}
\rho_{\rm SM}(a)\simeq \rho_{\rm SM}^{(0)}(a)~,\qquad\qquad
H(a)\simeq \sqrt{ \rho_{\rm SM}^{(0)}(a)/3M^2_{\rm pl}}\equiv
H^{(0)}(a)\;.
\end{equation}
The corrections to these expressions are doubly suppressed by the
product of $\alpha$ and the small ratio $\Sigma$ introduced in
eq.~(\ref{eq:Sigma}). As the main effect of DM on BBN is of order
$\alpha$, these corrections can be neglected. 

We now perturb the formula (\ref{Xnweak}) for the neutron abundance
after freeze-out. The perturbation function $\alpha\big(\phi(a)\big)$ 
enters through the Fermi constant
and the neutron-proton mass difference. To simplify the result, we assume
that during the relevant epoch temperature scales approximately
as
$T\propto 1/a$. This is equivalent to neglecting the change in the
  effective number of relativistic degrees of freedom, which is a good
  approximation as long as $T\gtrsim m_e$.
We arrive at,
\begin{equation} 
\label{XnWpert}
\begin{split}
\Delta X_{n,{\rm W}}=&\int_0^\infty \frac{da}{a} \cdot
\frac{m_{np}}{2T(1+\cosh{(m_{np}/T)})}\cdot
\exp\bigg(-\int_a^\infty \frac{da_1}{a_1} 
\tilde\lambda_{n\to p}(a_1)\bigg)
\\
&\times\bigg\{-\alpha(a) \tilde\lambda_{n\to p}(a)
+\int_a^\infty \frac{da_2}{a_2}\alpha(a_2)
\tilde\lambda_{n\to p}(a_2)
\bigg(4+\frac{m_{np}X_n^{\rm
  eq}}{T}-\frac{m_{np}J'}{TJ}\bigg)\bigg|_{a_2}\bigg\}\;,\\
\end{split}
\end{equation}
where $J'\equiv dJ(z)/dz$ and 
\begin{equation}
\label{tildelam}
\tilde\lambda_{n\to p}\equiv \frac{\lambda_{n\to p}}{H}
\Big(1+e^{-\frac{m_{np}}{T}}\Big)\;.
\end{equation}
As long as $\tilde\lambda_{n\to p} > 1$, i.e. the
reactions are faster than the 
expansion of the universe, the integrand is strongly suppressed
by the exponential factor. This is the case before the weak
freeze-out.
On the other hand, at low temperatures, the integral is cut off by the
hyperbolic cosine in the denominator. Thus, the integral is saturated
at temperatures around $T\sim T_W$. 
The two terms in the curly brackets can be traced back to the modification of the
equilibrium neutron abundance and the reaction rate respectively. Note that they enter with opposite signs. Depending on the behavior of
$\alpha(a)$, one or the other will win. 
This means that both over- and under-production of
neutrons is possible in different regions of  
ULDM parameters. This is in stark contrast with the approximation that the weak-interactions freeze-out instantaneously and $\Delta X_{n,W}\geq0$ for all ULDM parameters.

From the Saha equation (\ref{Saha}) we infer that the BBN temperature
scales as the deuterium binding energy. Hence,
\begin{equation}
\label{TBBNchange}
\fracchange{T_{\rm BBN}}\simeq \alpha\Big|_{\rm BBN}\;,
\end{equation}
and
\begin{equation}
\label{aBBNchange}
\fracchange{a_{\rm BBN}}\simeq -\fracchange{T_{\rm BBN}}\simeq
-\alpha\Big|_{\rm BBN}\;.
\end{equation}
This gives the perturbation of the
upper integration limit in the integral describing the neutron decay in
eq.~(\ref{He4 fraction}). The change of the lower limit is irrelevant.
Indeed, the ratio of the neutron decay width to the Hubble rate at
week freeze-out is $(\Gamma_n/H)_{\rm W}\sim 3\cdot 10^{-3}$. Thus, even an
order-one change
in $a_W$ would affect the helium abundance only at the level of a few
per mille. Of course, in our case the change is further suppressed by the
small coupling $\alpha$. Thus, we can set $a_{\rm W}$ in eq.~(\ref{He4
  fraction}) to be equal to its standard value $a_{\rm W}^{(0)}$.
The variation of the decay width $\Gamma_n$
takes place in the intervening times between the weak freeze-out and
the deuterium burning. Thus, the shift
$\Delta\Gamma_n$ must be evaluated at the appropriate scale
factor,
\begin{equation}
\fracchange{\Gamma_n}(a)=\alpha\big(\phi(a)\big)\;.
\end{equation} 
Combining everything together, we arrive at the following expression,
\begin{gather}
\fracchange{Y_{\rm p}} = 
\fracchange{X_{n,{\rm W}}}
+\frac{\Gamma_n^{(0)}}{H^{(0)}}\,\alpha\Big|_{\rm BBN}
-\Gamma_n^{(0)}\int_{a_W^{(0)}}^{a_{\rm BBN}^{(0)}}
\frac{da}{aH^{(0)}(a)}\,\alpha\big(\phi(a)\big)\;,
\label{MainAnalyticResult}
\end{gather}
where $\Delta X_{n,{\rm W}}$ is given by (\ref{XnWpert}).
Generally, we find that the weak freeze-out term dominates, however, there are regions of parameter space where delicate cancellations of take place, which leads to the non-trivial features of fig.~\ref{fig:results}. 
In the next subsection we will use this formula to derive the
constraints on the ULDM model. 

\paragraph{Jordan frame} It is instructive to repeat the above calculation
in the Jordan frame to highlight some subtleties in the connection between the two frames. Here, the effect of ULDM enters only via the modified
Hubble rate (see appendix~\ref{app:Einstein-Jordan map}),
\begin{equation}
\label{JordH}
\bar H(\bar a) = H^{(0)}(\bar a)\left[1 + \alpha(\bar a)\, \left(1+
  \frac{d\ln\alpha}{d\ln \bar a}\right)\right]\;,
\end{equation}
whereas all masses and widths remain unchanged. The perturbation of
the neutron abundance upon freeze-out, eq.~(\ref{Xnweak}), then reads,
\begin{equation}
\begin{split}
\label{XnWpertJ}
{\Delta \bar X}_{n,{\rm W}}=\int_0^\infty& \frac{d\bar a}{\bar a}\cdot
\frac{m_{np}}{2\bar T(1+\cosh{(m_{np}/{\bar T})})}\cdot
\exp\bigg(-\int_{\bar a}^\infty \frac{d\bar a_1}{\bar a_1} 
\tilde\lambda_{n\to p}(\bar a_1)\bigg)\\
&\times \int_{\bar a}^\infty\frac{d\bar a_2}{\bar a_2}\, \alpha(\bar a_2)
\bigg(1+\frac{d\ln \alpha}{d\ln \bar a_2}\bigg)
\tilde\lambda_{n\to p}(\bar a_2)\;.
\end{split}
\end{equation}
At first sight, this expression differs from the Einstein frame
expression (\ref{XnWpert}). In
particular, it
involves the derivative of $\alpha$ with respect to the scale
factor while the latter does not. However, it is straightforward
to verify that removing the derivative by
integration by parts and taking into account the dependence of the
temperature on the scale factor $T\propto 1/a$  
brings eq.~(\ref{XnWpertJ}) exactly to the form (\ref{XnWpert}).
While this equivalence is to be expected, it is worth mentioning that
an oversimplification of the problem could break it down. In
particular, this would be the case if, to estimate the neutron
abundance, one used the approximation of an
instantaneous
freeze-out at the temperature when the weak reaction
rate becomes equal to the Hubble rate. Then the Einstein frame
description would be sensitive only to the value of $\alpha$ at that moment,
whereas the Jordan frame one would also depend on its derivative.
Thus, consideration of the full kinetic equation (\ref{kinetic}) is
essential for the consistency of the analysis.

Similarly, the contribution accounting for neutron decay in
eq.~(\ref{He4 fraction}) in the Jordan frame is,
\begin{equation}
\begin{split}
&-\Delta\bigg(\int_{\bar a_{\rm W}}^{\bar a_{\rm BBN}}
\frac{d\bar a}{\bar a \bar H}\bar \Gamma_n\bigg)
=\Gamma_{n}^{(0)}\int_{a_{\rm W}^{(0)}}^{a^{(0)}_{\rm BBN}}
\frac{d\bar a_1}{\bar a_1 H^{(0)}(\bar a_1)}\;\alpha(\bar a_1)
\bigg(1+\frac{d\ln \alpha}{d\ln \bar a_1}\bigg)\\
&\qquad=
-\frac{\Gamma_n^{(0)}}{H^{(0)}}\alpha\bigg|_{\rm W}
+\frac{\Gamma_n^{(0)}}{H^{(0)}}\alpha\bigg|_{\rm BBN}
+\Gamma_n^{(0)}\int_{a_{\rm W}^{(0)}}^{a^{(0)}_{\rm BBN}}
\frac{d\bar a_1}{\bar a_1 H^{(0)}(\bar a_1)}\;\alpha(\bar a_1)
\bigg(1+\frac{d\ln H^{(0)}}{d\ln \bar a_1}\bigg)\;,
\end{split}
\end{equation}
where passing to the second line we have integrated by parts. As
discussed above, the first term is small and can be safely
neglected. For the derivative of the Hubble rate in the last term we
use eq.~(\ref{standCons}) from the appendix~\ref{app:Einstein-Jordan
  map}, 
which yields,
\begin{equation}
\frac{d\ln H^{(0)}}{d\ln \bar a}=-2+\frac{\Sigma}{2}\;.
\end{equation}
Neglecting, as usual, $\Sigma$-suppressed contributions, we again
reproduce the Einstein-frame result. This serves as a cross-check of
our calculation.

\subsubsection{Analytic constraints from Helium-4}

\begin{figure}[t]
\centering
	\includegraphics[width=0.49\textwidth]{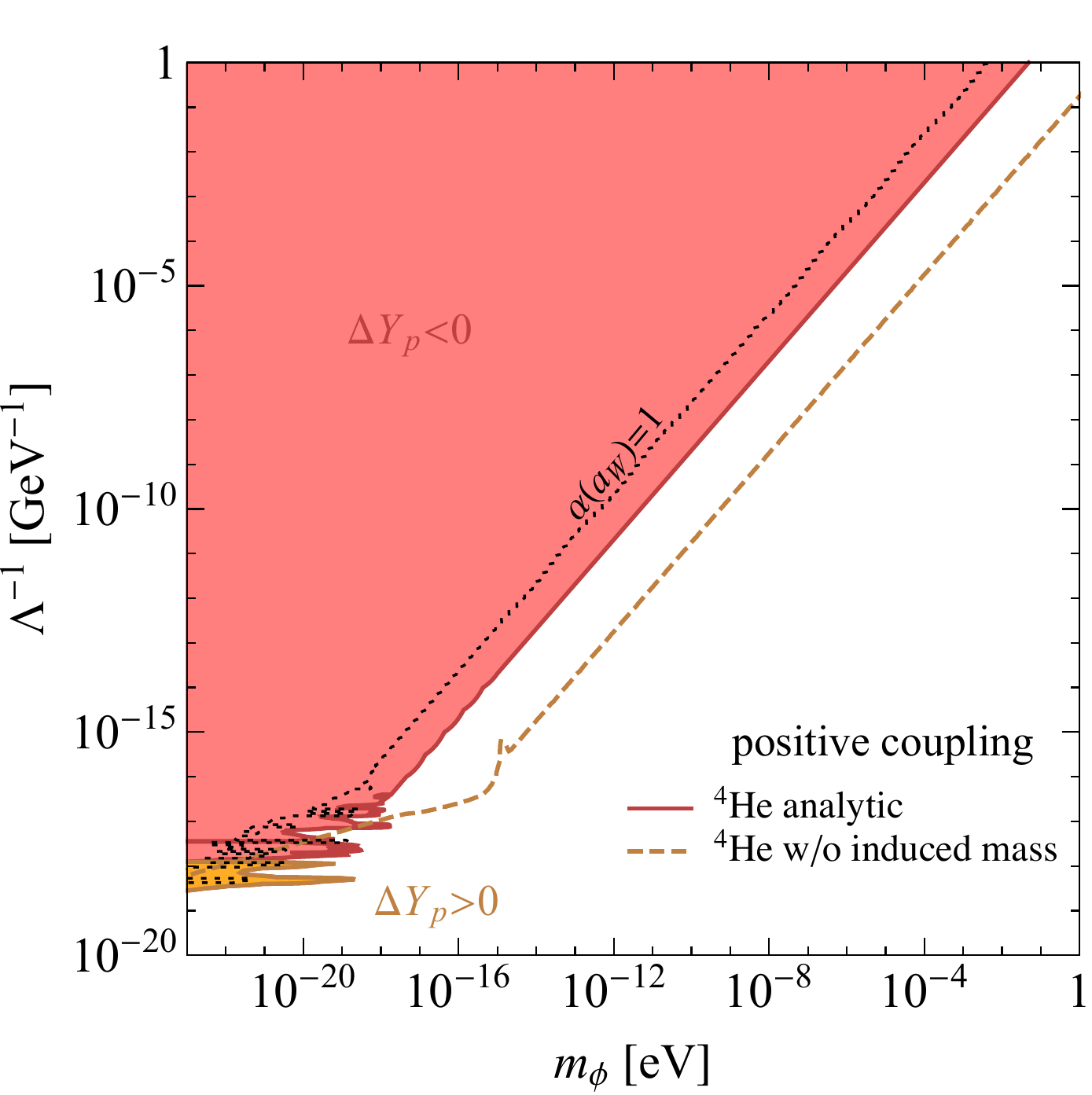}~~
	\includegraphics[width=0.49\textwidth]{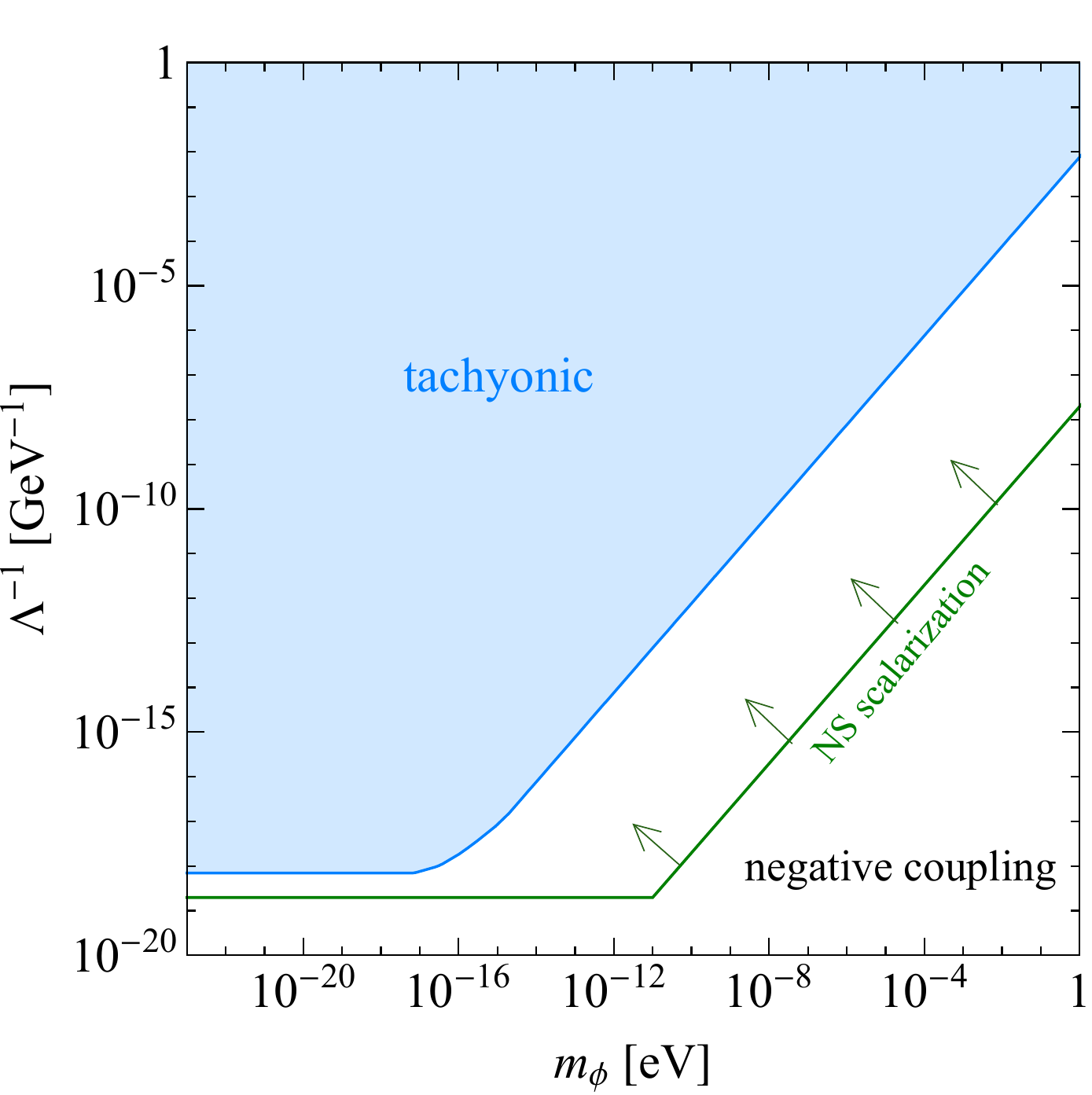}
	\caption{\small {\bf Left:} BBN constraints on the scalar dark
 matter parameter space in the case of positive coupling  
$\alpha=+\phi^2/\Lambda^2$. The shaded region is excluded at 
95\%CL. The red shading corresponds to
    underproduction of $^4$He while the orange shading
 corresponds to overproduction. Orange dashed line shows the
 constraint obtained in an analysis that neglects the induced
dark matter mass and uses instantaneous weak freeze-out
 approximation. For the parameters to the left of the black
dotted line the coupling $\alpha$ becomes non-perturbative
 at the time of weak
freeze-out (scale factor $a_W=10^{-9.6}$). {\bf Right:} Parameter space
 of the model with negative coupling
 $\alpha=-\phi^2/\Lambda^2$. The blue shaded region
corresponds to tachyonic instability during BBN. It is
excluded unless the initial conditions for dark matter are
 extremely fine-tuned. The region above the green line admits
 spontaneous scalarization of neutron stars.} 
\label{fig:results}
\end{figure}

 By combining the limits on the $^4$He abundance (\ref{He4
  constraint}) with the impact on BBN derived in the previous
section (\ref{MainAnalyticResult}), we arrive at the constraints for
ultralight scalar DM. 
We first consider the case of an ULDM field with positive coupling
$\alpha=\phi^2/\Lambda^2$. This yields constraints on the DM
parameter space shown in the left panel of
fig.~\ref{fig:results}. Also shown are the constraints from a model
which neglects the effects due to the DM induced mass and assumes
an instantaneous weak freeze-out~\cite{Stadnik:2015kia}. 

Including the DM induced mass dramatically modifies the
constraints. They are strengthened at low masses $m_\phi\lesssim
10^{-22}~{\rm eV}$, weakened by up to two orders of magnitude at high
masses $m_\phi\gtrsim 10^{-18}~{\rm eV}$, and show non-trivial features
at intermediate masses. To understand this behavior, we first look at
the two limiting cases at low and high mass where the ULDM field, and
therefore $\alpha$, has a straightforward time-dependence. For very
light masses, the field is 
in the Hubble-friction (H) dominated phase during BBN\footnote{For the
purposes of this discussion we consider neutron freeze-out as part of
the BBN epoch.} where
$H\gg m_{\phi,{\rm eff}}$.
Thus, the field is frozen at a constant value and its effect is
analogous to changing $M_{\rm pl}$ as shown in
eq.~\ref{Mplvar}. Later on the field begins oscillating with decreasing
amplitude. The induced mass shifts the start of oscillations to an
earlier time, which gives more room for the amplitude to decay. This
increases the frozen value of the field at BBN and leads to stronger
constraints. 

On the other extreme, for high masses
($m_\phi \gtrsim 10^{-18}$ eV) the ULDM field is dominated by the
induced mass (I-regime) throughout BBN. Here the field is rapidly
oscillating and its amplitude is described by the WKB formula
(eq.~\ref{WKBampl}) both at BBN and at later times. 
The effective mass in the denominator strongly reduces the
$\phi$-amplitude at BBN, compared to the case without induced mass,
weakening the constraints. Matching the WKB amplitude to the
present-day DM density $\rho_{{\rm DM},0}$ and using that today
$m_{\phi,{\rm eff}}= m_\phi$, whereas at BBN it is 
$m_{\phi,{\rm eff}}= \sqrt{2\Theta_{\rm SM}}/\Lambda$, we can find
$\alpha$ as a function of the scale factor at BBN,
\begin{equation}
\alpha(a)=\frac{1}{m_\phi\Lambda}\cdot\frac{\rho_{{\rm
      DM},0}}{a^3\sqrt{2\Theta_{\rm SM}(a)}}\;.
\end{equation} 
This expression implies that BBN is sensitive to the
combination $m_\phi\Lambda$, as indeed seen from
fig.~\ref{fig:results}. Notice that in this regime the field amplitude
features a rapid decrease with the scale factor. Thus most of the
constraining power comes from the earliest stage of BBN, i.e. the weak
freeze-out. 

The scaling in the intermediate mass range is more
complicated due to the non-trivial time dependence of the induced
mass. 
In this parameter region the field exhibits oscillations on the time
scale comparable to the expansion rate of the universe. Depending on
whether the oscillations happen to be on the peak (in the trough)
during the relevant stages of BBN, the constraints are strengthened
(weakened). This explains the oscillating features in the exclusion
line clearly visible in fig.~\ref{fig:results}.

The reduction of BBN sensitivity due to the rapid time-dependence of the DM
field allows the coupling $\alpha$ to approach order-one values during
BBN. This is dangerous, since in our analysis we assume $\alpha$ to be
small and expand at linear order in it. As a criterion for the
validity of this approximation we impose the requirement that $\alpha$
should be less than 1 at the moment of the weak freeze-out marking the
beginning of BBN. The corresponding parameter region is delimited by
the black dotted line in fig.~\ref{fig:results}. We have verified that
our constraints always lie in the region where this criterion is
satisfied. Still, our tests show that inclusion of non-linear terms in
$\alpha$ can shift the constraints on $m_\phi$ or $\Lambda$ by a
factor of~2. 

The dynamics of neutron freeze-out has an important qualitative impact on
the predictions of the model.  
From eq.~\ref{XnWpert} we saw that, depending on the choice
of ULDM parameters, neutrons, and therefore $^4$He, can be 
either over- or under-produced. In the H-dominated regime, the modification of the reaction
rates dominates and we increase the amount of $^4$He. In the
I-dominated regime, the modification of the equilibrium neutron abundance
dominates and the amount of $^4$He decreases. The transition between
over- and under-production happens along the horizontal line 
$\Lambda^{-1}\approx 10^{-18}~{\rm GeV}^{-1}$ in fig.~\ref{fig:results}.
Note that the instantaneous weak
freeze-out approximation leads to an over-production of $^4$He across
all of the parameter space, in stark contrast to the actual predictions.  

For a negative coupling $\alpha= -\phi^2/\Lambda^2 $
DM becomes tachyonic in the induced mass dominated phase. This
corresponds to the blue shaded region in the right panel of
fig.~\ref{fig:results}.
In this case the initial conditions for the field $\phi$ must be
strongly fine-tuned to avoid its overproduction. We find this
fine-tuning unappealing and do not explore the tachyonic regime
further. The ULDM does not produce any observable effects on BBN
outside of the tachyonic region. 
Interestingly, the allowed parameter space contains
a band where the DM stays non-tachyonic during
cosmological evolution since the start of BBN, but still has a coupling strong
enough to develop instability inside 
extremely compact objects
such as neutron stars. This may lead to a phenomenon known as
scalarization (see
e.g.~\cite{Damour:1993hw,Damour:1996ke,Harada:1997mr,Pani:2010vc})
during which the compact object spontaneously acquires a scalar charge,
and can have observable signatures in {\it e.g.} binary pulsar
systems. The criterion for acquiring a scalar charge is that
$2\Theta_{\rm NS}/\Lambda^2\gtrsim \max\{m_\phi^2, R^{-2}\}$, where
$\Theta_{\rm NS}$ is the trace of the neutron star EMT and
$R\sim 10\,{\rm km}$ is the neutron star size. Neglecting for an
estimate the pressure contribution and taking 
the neutron star density $ \rho_{\rm NS} \simeq 3
\times 10^{14}~\rm{g/cm}^3 $~\cite{Lattimer:2012nd}, we obtain that
scalarization is possible in the region above the green line in
fig.~\ref{fig:results}. A more detailed investigation of scalarization
is justified, but is beyond the scope of this work. 
\subsection{Numerical constraints from Helium-4 and Deuterium }
\label{sec:Jordan}

The calculation of the abundances of the light elements beyond $^4$He requires the study of a complex network of nuclear interactions. Traditionally, this problem is handled with numerical codes such as {\tt AlterBBN}~\cite{Arbey:2018zfh}, {\tt PRIMAT}~\cite{Pitrou:2018cgg} and {\tt PArthENoPE}~\cite{Consiglio:2017pot}. 
Implementing our model into such a code in the Einstein frame involves continuously changing fermion masses and reaction rates, which would take a substantial effort. Instead of undertaking such an effort, we make use of the Jordan frame where the SM is left unchanged but instead evolves in a universe with modified expansion rate (eq.~\ref{JordH}).
In this numerical study we restrict to the case of positive DM coupling, $\alpha>0$.

We use the numerical code {\tt AlterBBN}~\cite{Arbey:2018zfh}, which parameterizes changes to the evolution of the scale factor (and Hubble parameter) by introducing a dark density component in the Friedman equation. The dark density required to reproduce the Hubble rate (eq.~\ref{JordH}) has the form,
\begin{equation}
\label{rhodark}
    \rho_{\rm dark}=6M_{\rm pl}^2 \big[H^{(0)}(\bar a)\big]^2 \alpha(\bar a) \left(1 +  \frac{d \ln \alpha}{d \ln \bar a} \right).
\end{equation}
Note that, since $\alpha$ can be rapidly decreasing or oscillating, the derivative term in brackets can be negative, pushing $\rho_{\rm dark}<0$. 
Using the standard thermal history of the universe, we convert $\rho_{\rm dark}$ into a function of temperature. Then
we import it into {\tt AlterBBN v2.2} through the function {\tt Init\_dark\_density\_table}, such that the evolution is modified to correspond to that implied by our model. In this way one can make numerical estimates of not only $^4$He but also D, $^3$He, $^6$Li, $^7$Li, and $^8$Be. Unfortunately, experimental data suitable for comparison with these predictions only exist for the elements $^4$He, D, and the combined lithium abundance~\cite{Tanabashi:2018oca}.

One needs to proceed with caution in defining $\rho_{\rm dark}$. 
We find that in some regions of parameter space the DM amplitude may become large ($\phi^2/\Lambda^2 \sim\mathcal{O}(1)$) in the time range referenced by {\tt AlterBBN}. Here $\rho_{\rm dark}$, calculated according to (eq.~\ref{rhodark}), can exceed the standard model density. This is inconsistent because it signals the breakdown of the linear expansion in $\alpha$ used in our analysis. 
To avoid this problem, we implement a cut which limits the magnitude of dark density to half of the SM density, i.e. we require that the dark density is bounded by $|\rho_{\rm dark}|<0.5\rho_{\rm SM}$. This cut, which is only necessary in the region near the breakdown of the model, can be understood as a conservative measure, which discards constraints arising from  $ \alpha\sim\mathcal{O}(1)$. We only report constraints where $\alpha <1$ at the time of the weak freeze-out. However, {\tt AlterBBN} begins integrating at earlier times, where $\alpha$ may be larger than at weak freeze-out. Thus, the cut is used for regions where $\alpha$ is small at freeze-out but grows to $\mathcal{O}(1)$ towards the earliest times probed by {\tt AlterBBN}. 

\begin{figure}
	\centering
	\includegraphics[width=0.48\textwidth]{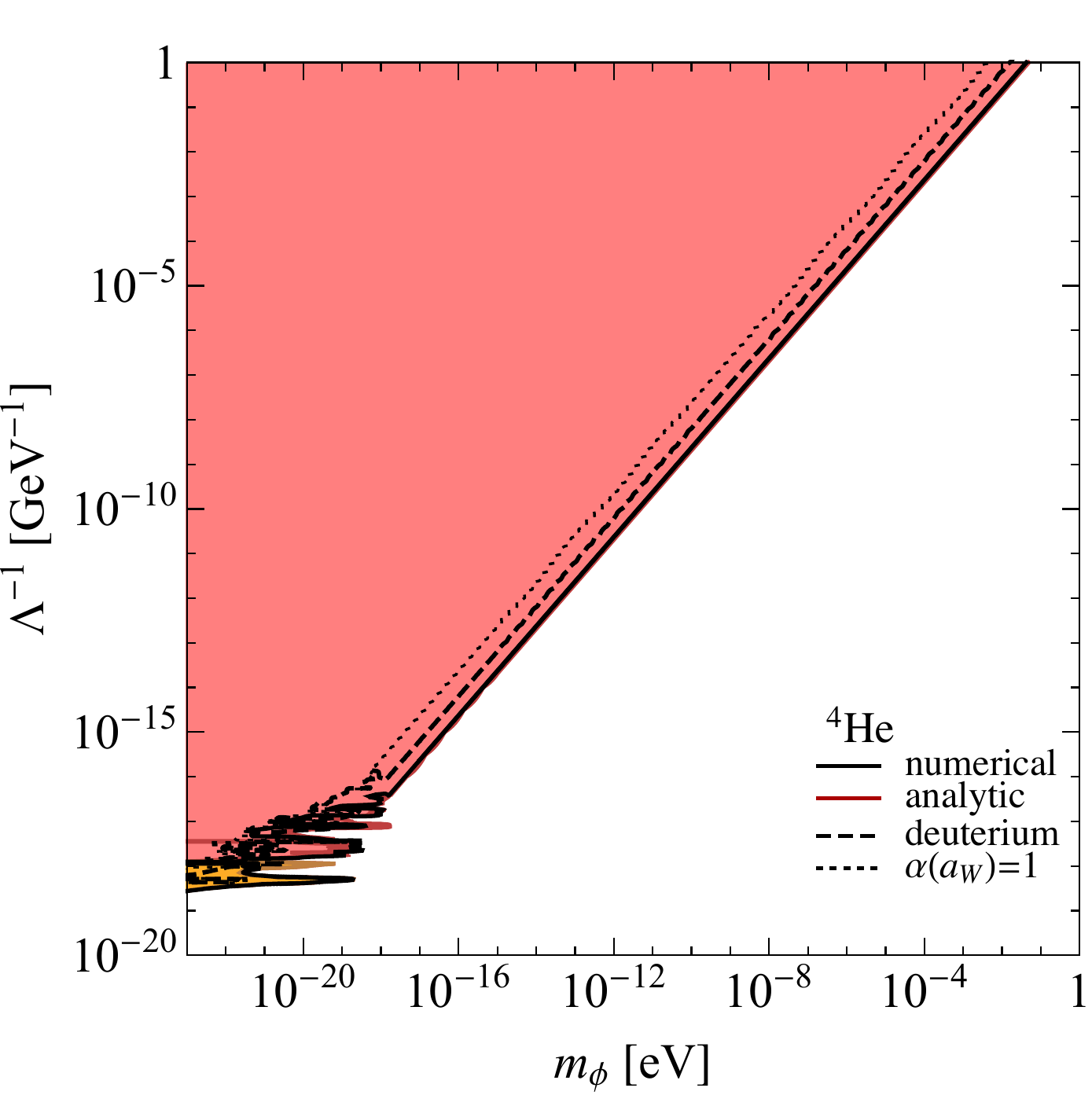}~~~~
	\includegraphics[width=0.48\textwidth]{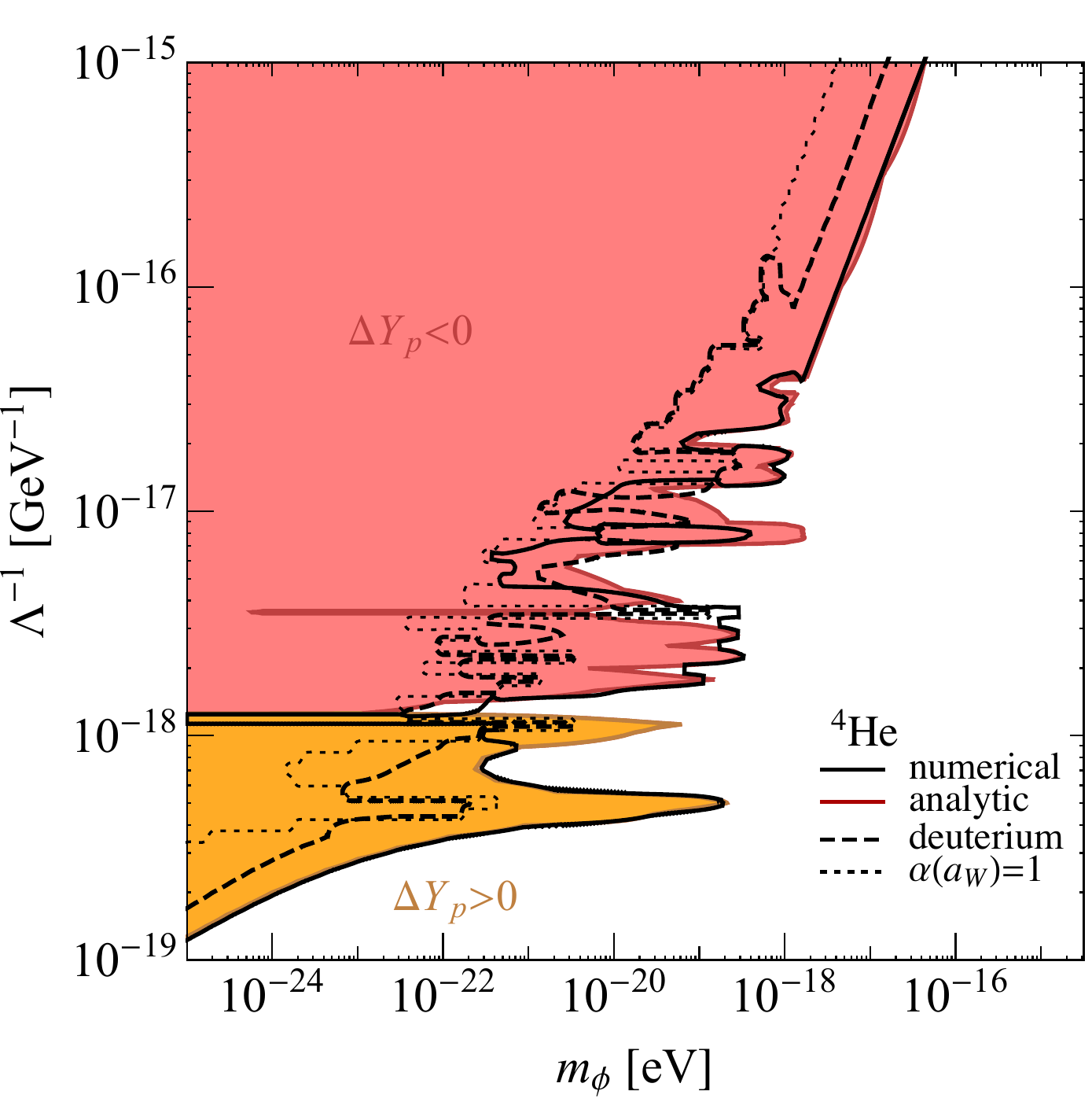}
	\caption{\small The numerical constraints on $^4$He (black, solid) and D (black, dashed) derived using {\tt AlterBBN} as compared to the constraints from the analytical approximation of the preceding section (red and orange regions). Again, red corresponds to underproduction of $^4$He and orange corresponds to its overproduction. Deuterium tends to be  overproduced.
	The black dotted line delimits the region of parameters where the coupling becomes large at the scale factor $a_W=10^{-9.6}$ corresponding to the weak freeze-out, $\alpha(a_W)>1$.
	To the left of this line higher-order terms in the Taylor expansion of the function $\alpha(\phi)$ become important. 
	The right panel zooms in to the region where the constraints exhibit oscillations. Here, the constraints should be taken with caution as they are sensitive to non-linear effects in $\alpha$.}
	\label{fig:resultsAlter}
\end{figure}

The numerical $^4$He constraint agrees well with the analytical counterpart in the vast majority of parameter space, as seen in fig.~\ref{fig:resultsAlter}. The minor discrepancies in the oscillating region can be traced back to the cuts, which are required in the Jordan frame analysis as described above, but are absent in the analytical treatment. The cuts have a particularly large effect in the oscillating region, because this region features a pronounced degree of cancellation, such that larger values of $\alpha$ are probed. This also implies that the oscillating region is more sensitive to non-linearities in $\alpha$ and 
model-specific higher-order terms in the Taylor expansion of the function $\alpha(\phi)$.
These are not captured in our analysis, and so the constraints in this region should be interpreted with care. There is no analytical counterpart for the numerical D exclusion, which is shown in the same figure. Note that the $^4$He constraint is dominant for most of parameter space.

With regards to lithium, it is well known that predictions of standard BBN theory do not match the observed abundances. This is also present in our model for the regions of parameter space which are compatible with the observed $^4$He and D abundances. Still, it is worth noting that the DM modification tends to slightly reduce lithium production, but this change is vastly insufficient to resolve the lithium problem.

\section{Additional constraints}
\label{sec:addconstraints}
The quadratic coupling of a new scalar particle to the SM is subject to many other astrophysical constraints,  ranging from the anomalous cooling of supernovae, modifications to the timing of binary pulsar systems, and black hole superradiance, as well as laboratory constraints coming from atomic clocks and torsion pendulums. 
The constraints from torsion balance experiments and atomic clocks probe WEP violation, and thus do not apply to our model. 
In what follows, we briefly summarize the strongest constraints on the quadratic coupling of a new scalar particle.

{\bf Supernova cooling:} New light particles with masses below the average temperature of the core of a supernova, $T_{\rm core}^{SN}\simeq 30$ MeV, can be produced within the supernova core and free-stream out, thus removing energy from the supernova. 
This new channel for energy loss can alter the  supernova neutrino luminosity.
Although a proper calculation of the free-streaming process is involved and requires complicated simulations, one can approximate an upper limit on the instantaneous luminosity of new particles produced in the supernova based on the observation of SN1987a~\cite{Raffelt:1990yz}:
$L_{\rm new}\leq L_\nu\simeq3\times 10^{52}{~\rm erg/s}$. If the instantaneous luminosity of new particles exceeds this value when the core reaches peak density, $\rho_c\simeq 3\times 10^{14}{~\rm g}/{\rm cm}^3$ and temperature $T_c\simeq 30$ MeV, then the energy spectrum of the neutrino burst from SN1987a becomes inconsistent with observations. This is known as the ``Raffelt criterion".\footnote{However, the robustness of this criterion has been called into question~\cite{Bar:2019ifz}.}
The dominant production process for the scalar is the bremsstrahlung of neutrons $nn\to nn\phi\phi$ due to the high density of neutrons in the proto-neutron star. The rate of $\phi$ production can be estimated as~\cite{Olive:2007aj}
\begin{equation}
    \Gamma_{nn\to nn\phi\phi}\simeq\sigma_{nn}\times\frac{\rho_c^2 T_c^{7/2}}{12\pi^4\Lambda^4 m_n^{1/2}}\, ,
\end{equation}
where $m_n$ is the neutron mass, and
where $\sigma_{nn}\simeq 25$ mb~\cite{Hanhart:2000er} is the elastic nucleon cross-section. 
The Raffelt criterion can be rewritten as $\Gamma_{nn\to  nn\phi\phi}\lesssim 10^{-14}{~\rm MeV}^5$, which corresponds to $\Lambda\gtrsim 14.5$ TeV.

{\bf Fifth-force experiments:} 
 Fifth-force searches traditionally set some of the strongest constraints on ultra-light scalar particles~\cite{Kapner:2006si,Adelberger:2006dh}. The quadratic coupling of the scalar produces a fifth-force at leading order through the exchange of a pair of $\phi$ between two fermions. This generates a potential of the form 
\begin{equation}
    V(r)=\frac{1}{r^3}\frac{m_f^2}{64\pi^3 \Lambda^4},
\end{equation}
in contrast to the usual $1/r$ Yukawa potential from linear-couplings. As a consequence, the constraints from fifth-force experiments on the quadratic coupling are much weaker than those on the linear coupling~\cite{Kapner:2006si,Adelberger:2006dh}. 
Limits on the deviation from the typical $1/r$ behavior of the gravitational potential require $\Lambda\gtrsim2$ TeV for $m_\phi\lesssim 10^{-4}$ eV~\cite{Olive:2007aj}. 
Note, however, that the presence of a non-zero $\phi$ background induces an effective linear, time-dependent coupling between $\phi$ and matter.
It would be interesting to reinterpret existing constraints in the context of this effect~\cite{Hees:2018fpg}.\footnote{We thank David E. Kaplan and Yevgeny Stadnik for discussion of this point.}

{\bf Galaxy formation and Ly-$\alpha$:} 
A lower limit on the scalar DM mass comes from the cosmological structure formation. The latter becomes affected if the de Broglie wavelength of the scalar is larger of comparable to the relevant astrophysical scales, such as the size of dwarf galaxies $R\sim 1$ kpc. Requiring that modifications of the structure formation are compatible with observations leads to the bound $m_\phi\gtrsim 10^{-22}$~eV~\cite{Schive:2014dra,Bozek:2014uqa,Hlozek:2014lca}, assuming that the scalar composes the entirety of DM density. 
 A more stringent lower bound comes from the measurement of the Ly-$\alpha$ forest, which requires $m_\phi\gtrsim 10^{-21}$ eV~\cite{Irsic:2017yje,Kobayashi:2017jcf}. 
 This is supported by complementary analyses of the galactic rotation curves~\cite{Bar:2018acw} and 
 the subhalo mass function~\cite{Schutz:2020jox}. If the detection of a global 21-cm absorption signal~\cite{Bowman:2018yin} is confirmed, it will set a similar bounds~\cite{Schneider:2018xba,Lidz:2018fqo,Sullivan:2018szg}.
 In addition, the fluctuations from light scalar DM can heat up the cores of galaxies and modify their dynamics. Based on the properties of the star cluster in Eridanus II, ref.~\cite{Marsh:2018zyw} has inferred the bound $m_\phi\gtrsim 10^{-19}$ eV. All these limits can be relaxed if the scalar is only a sub-component of the dark matter.

{\bf Pulsar Timing and Stochastic Gravitational Waves:} 
Binary pulsars are systems whose dynamics are measured with exquisite precision. As a result, they are highly sensitive to any new physics that changes their dynamics~\cite{Manchester:2015mda,Kramer:2016kwa}.
As discussed, the addition of an ultralight scalar field with universal couplings will perturb the masses of the SM particles, and thus change the masses of the stars in the binary, resulting in a change in the orbital period. The effect is especially pronounced if the frequency of the perturbation matches the harmonics of the binary orbital frequency. In this situation, the changes in orbital parameters are resonantly amplified and can lead to measurable effects~\cite{Blas:2016ddr,Blas:2019hxz}. Current data give strong bounds on the coupling in the resonant bands of masses; future observations with the Square Kilometer Array telescope will significantly strengthen the constraints and extends the coverage.
Complementary constraints are provided by Pulsar Timing Arrays (PTA)~\cite{Porayko:2014rfa} as well as searches for stochastic gravitational waves by the Cassini (CAS) space mission~\cite{Armstrong:2003ay}. These constraints are weaker than the BBN constraints derived in this work. 

\begin{figure*}[t]
\centering
\includegraphics[width=0.9\textwidth]{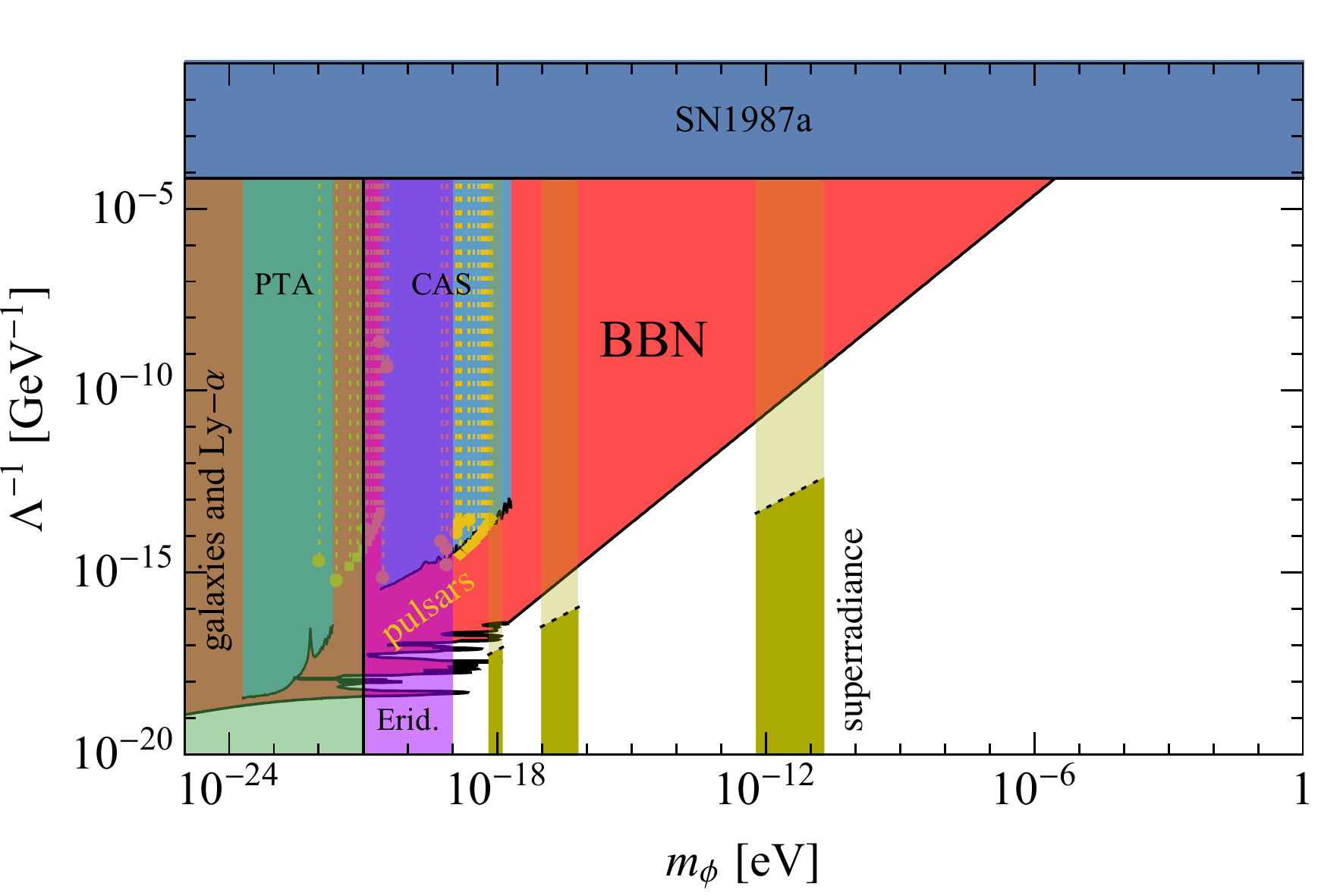}
\caption{\small Summary of constraints on the scale of the universal quadratic coupling, $1/\Lambda$, as a function of scalar mass $m_\phi$. Only the case of positive coupling is shown. The bounds from the $^4\rm{He}$ and D abundances are shown in the red shaded region. 
Additional constraints come from supernova cooling and fifth-force searches (blue),  superradiance (yellow), the deBroglie wavelength of the smallest dwarf galaxies along with bounds from Ly-$\alpha$ measurements (green), and Eridanus II (purple). Above the black dotted lines, the induced mass from the black hole accretion disk exceeds $m_\phi$ and the dynamics of superradiance may be affected. The constraints from measurements of the binary pulsar orbital period are given in the yellow dots, corresponding to the resonant bands. Also shown are constraints inferred from the bounds on stochastic gravitational waves by Cassini (CAS) and Pulsar Timing Arrays (PTA). See text for more detail.
\label{fig:finalconstraints}}
\end{figure*}

{\bf Black Hole Superradiance:} 
Ultralight scalar fields can form gravitationally bound states with black holes if their Compton wavelength is of comparable size to the Schwarz\-schild radius of the black hole. The scalar fields extract angular momentum from the black hole through a process known as superradiance, which cause rapidly spinning black holes to spin down~\cite{Arvanitaki:2009fg,Arvanitaki:2010sy,Brito:2015oca}.\footnote{A similar phenomenon can also occur with millisecond pulsars if the scalar has a Yukawa-type coupling to neutrons~\cite{Kaplan:2019ako}.} Observations of old, near-extremal black holes can therefore exclude the existence of weakly-interacting scalars in the mass ranges of
$m_\phi\in [10^{-18.2},10^{-17.6}]$, $ [ 10^{-16.7},10^{-16.1}],$ and $[10^{-13},10^{-10.8}]$ eV~\cite{Arvanitaki:2014wva}.
Measurements of M87$^*$ by the Event Horizon Telescope~\cite{Akiyama:2019cqa} further constrain scalar masses $m_\phi\in[2.9\times10^{-21}, 4.6\times 10^{-21}]$ eV~\cite{Davoudiasl:2019nlo}. Note that the accretion disk around black holes will modify the effective mass of the scalar DM, which in turn will affect the dynamics of superradiance. We estimate the effect of the accretion disk by comparing the induced mass due to the baryonic density of the accretion disk~\cite{Abramowicz:2011xu} to the bare mass $m_\phi$.  We denote the value of $\Lambda$ at which the induced mass is comparable to $m_\phi$ by a black dotted line in fig.~\ref{fig:finalconstraints}, above which the superradiance bounds may change. A more sophisticated statistical analysis, taking into account the ensemble of black hole spin measurements, can increase the range of masses constrained~\cite{Stott:2018opm}.  
These bounds assume that the self-interactions of the scalar are negligible. Large self-interactions and other non-linear effects can render superradiance ineffective, thus allowing for the scalar to evade the aforementioned bounds~\cite{Arvanitaki:2010sy,Fukuda:2019ewf,Mathur:2020aqv}. 

A summary of all the constraints described in this section, along with the BBN constraints from this work, are collected in fig.~\ref{fig:finalconstraints}. 
%
\section{Conclusions}
\label{sec:conclusions}
In this work, we have investigated the effect of ultralight scalar DM with universal quadratic couplings to SM fields on predictions of BBN.
This type of coupling preserves the weak equivalence principle and does not give rise to a long-range force, thereby evading numerous laboratory tests. We took into account the full dynamics of BBN, as well as DM cosmological evolution. 
We have found that the precision measurements of the primordial $^4$He abundance constrain a large portion of ULDM parameter space as shown in fig.~\ref{fig:finalconstraints}. 

The quadratic coupling has an important effect of the evolution of DM endowing it with an effective mass proportional to the energy density of the SM environment. This leads to substantial corrections to previous calculations~\cite{Stadnik:2015kia}, where this effect was neglected. Specifically, for positive coupling the constraints from the full ULDM evolution are about two orders of magnitude weaker for $m_\phi\gtrsim 10^{-18}$ eV and have non-trivial features at $m_\phi\lesssim10^{-18}$ eV where the ULDM exhibits oscillations on a time scale comparable to the expansion rate of the universe. In contrast, at very low masses our constraints are stronger. For negative coupling we find the DM has a tachyonic instability in a large portion of the parameter space, requiring an extreme fine-tuning of initial conditions to avoid its overproduction.

In addition, we moved beyond the instantaneous approximation for neutron freeze-out and used the full kinetic description. We have found that it qualitatively changes the result for the $^4$He abundance, which is predicted to decrease in much of the parameter space; 
in contrast, the instantaneous approximation erroneously predicts an over-production of $^4$He across all of parameter space.   

 The universal coupling studied in this work allows for treatment in an alternative frame, the Jordan frame, where the ULDM modifies the metric, but does not couple directly to the matter fields.
 The modified metric manifests itself as a modification to the Hubble parameter, which we implemented into a numerical code {\tt AlterBBN v2.2}. The numerical analysis allows us to determine not only $^4$He abundance, but also that of other light elements, such as deuterium and lithium. We found the limits based on deuterium are subdominant to $^4$He throughout the parameter space.
  Curiously, the ULDM leads to a slight reduction of the Li abundance, although this is vastly insufficient to resolve the ``lithium problem". 

Our results show that BBN sets the strongest constraint on the ULDM coupling for the majority of the parameter space where $10^{-19}\lesssim m_\phi\lesssim 10^{-6}$ eV. However, we note a few caveats. 
Our constraints are subject to the assumption that SM and ULDM are described by a simple Lagrangian (eq.~\ref{EinsteinLag}) all the way back to BBN. We further assumed that we can Taylor expand the coupling $\alpha(\phi)$ keeping only the leading-order term. Violation of these assumptions may modify the constraints. Inclusion of a self-interaction of the DM field or its more complicated coupling to SM can significantly alter 
 the dynamics of the ULDM and lead to rich phenomenology (see {\it e.g.}~\cite{Tkachev:1986tr,Goodman:2000tg,Peebles:2000yy,Arvanitaki:2014faa,Fan:2016rda,Belokon:2018hrn}). The specific form of these non-linear terms could be determined within a more complete setting describing the UV origin of the universal coupling. 
  Further study along these lines is warranted and we leave this for future work. 

\acknowledgments
We thank Asimina Arvanitaki, Diego Blas, Mark Hertzberg, Junwu Huang, and Anna Tokareva for clarifying discussions. 
We are grateful to Alexandre Arbey for his advice on the use of the {\tt AlterBBN} package.
We also thank Prateek Agrawal, Diego Blas, and Yevgeny Stadnik for useful comments on an earlier version of this manuscript. PS is grateful to CERN for hospitality during the Summer Student Programme, where this study was initiated. TTY thanks the participants of the 2019 PACIFIC Conference and the Aspen Workshop ``Progress after Impasse: New Frontiers in Dark Matter" for numerous enlightening discussions, as well as the hospitality of the Aspen Center for Physics which is supported by National Science Foundation grant PHY-1607611, where part of this work was completed. The work of SS has been supported by the Tomalla Foundation. PS is supported by the Deutsche Forschungsgemeinschaft under Germany’s Excellence Strategy - EXC 2121 “Quantum Universe” - 390833306.

\begin{appendices}

\section{Variations of dimensionful parameters}
\label{app:variations}

In sec.~\ref{sec:scalar} we have shown that in the Einstein frame the
masses of the fundamental particles, such as leptons, quarks and
vector bosons, are all rescaled by the universal interaction in
the same way (\ref{massresc}).
We argued also that preservation of the
weak equivalence principle requires $\Lambda_{\rm QCD}$ to obey the
same law (\ref{LQCDvar}). More generally, WEP implies that
the dimensionless ratios of all SM couplings are independent of $\phi$. Then the
dimensionful quantities should scale according to their mass
dimensions. It is instructive to see in some detail how it works out
for the quantities controlling the BBN dynamics.

Consider first the Fermi constant $G_F$. It is related to the $SU(2)$
coupling $g$ and the $W$-boson mass as
\begin{equation}
G_F=\frac{g^2}{4\sqrt{2}M_W^2}\;.
\end{equation}
The coupling $g$ is dimensionless and not affected by
ULDM.\footnote{Upon inclusion of quantum corrections, this statement holds for
  $g(\mu)$ if the renormalization group scale $\mu$ is defined in
  terms of the physical particle masses.}
Thus, the variation of $G_F$ comes exclusively from the variation of
$M_W$,
\begin{equation}
\frac{\Delta G_F}{G_F}=-2\alpha(\phi),
\end{equation}
assuming $\alpha(\phi)$ is small.

The neutron-proton mass difference $m_{np}$ receives
contributions from the isospin symmetry breaking by the quark masses and
from the electromagnetic interaction,
\begin{equation}
m_{np}=\delta_{\rm iso}\frac{(m_d-m_u)}{(m_d-m_u)^{(0)}}
+ \delta_{\rm em}\frac{\alpha_{\rm em}\Lambda_{\rm QCD}}{(\alpha_{\rm em}\Lambda_{\rm QCD})^{(0)}}\, ,
\end{equation}
where
$\delta_{\rm iso}=2.05\pm 0.30$~MeV and  $\delta_{\rm em}=
-0.76\pm0.30$~MeV~\cite{Gasser:1982ap}. 
Here $\alpha_{\rm em}$ is the fine structure constant.
A more recent {\it ab initio} 
lattice calculation finds\footnote{The quoted uncertainties are squared
sums of the statistical and systematic errors.}
$\delta_{\rm iso}=2.52\pm 0.29$~MeV and
$\delta_{\rm em}= -1.00\pm
0.17$~MeV~\cite{Borsanyi:2014jba}. Independently of the precise
values, we have 
\begin{equation}
   \frac{\Delta m_{np}}{m_{np}}=\frac{\delta_{\rm iso}}{\delta_{\rm
       iso}+\delta_{\rm em}}\frac{\Delta(m_d-m_u)}{m_d-m_u}+
\frac{\delta_{\rm em}}{\delta_{\rm iso}+\delta_{\rm em}}
\frac{\Delta\Lambda_{\rm QCD}}{\Lambda_{\rm QCD}}\simeq \alpha(\phi)\, ,
\end{equation}
where we have used eqs.~(\ref{massresc}), (\ref{LQCDvar}).

The neutron decay rate can be approximated by~\cite{Coc:2006sx}
\begin{gather}
\Gamma_n= \frac{1+3g_A^2}{2\pi^3} G_F^2 m_e^5\, P(m_{np}/m_e)\, ,
\end{gather}
where $g_A\simeq 1.26$ is the nucleon axial charge and
\begin{equation}
P(x) = \frac{1}{60}\Big[\left( 2x^4 -9x^2-8 \right)\sqrt{x^2-1}
+15 x\ln\left(x+\sqrt{x^2-1}\right)\Big]\, .
\end{equation}
Taking the variation of all masses as before we arrive at the
fractional change in decay rate 
\begin{equation}
\frac{\Delta \Gamma_n}{\Gamma_n} = 5 \fracchange{m_e} + 2
\fracchange{G_F}+ 
\frac{P'}{P}\frac{m_{np}}{m_e}\left( \fracchange{m_{np}}- \fracchange{m_e} \right)
\simeq \alpha(\phi)\, .\label{gamma perturbation}
\end{equation}
The sensitivity of the deuterium binding energy 
$B_D$ can be estimated in a variety of ways~\cite{Uzan:2010pm}.
 Our results are not particularly sensitive to the exact scheme. Following~\cite{Coc:2006sx}, we take  
\begin{equation}
 \fracchange{B_D} \simeq 18\fracchange{\Lambda_{\rm
    QCD}}-17\fracchange{m_q}\simeq \alpha(\phi)\;, 
\end{equation}
where $\Delta m_q/m_q$ is the variation of the quark masses.

\section{Dark matter evolution} \label{app:DMevolution}
\begin{figure}[t!]
\centering
\includegraphics[width=\textwidth]{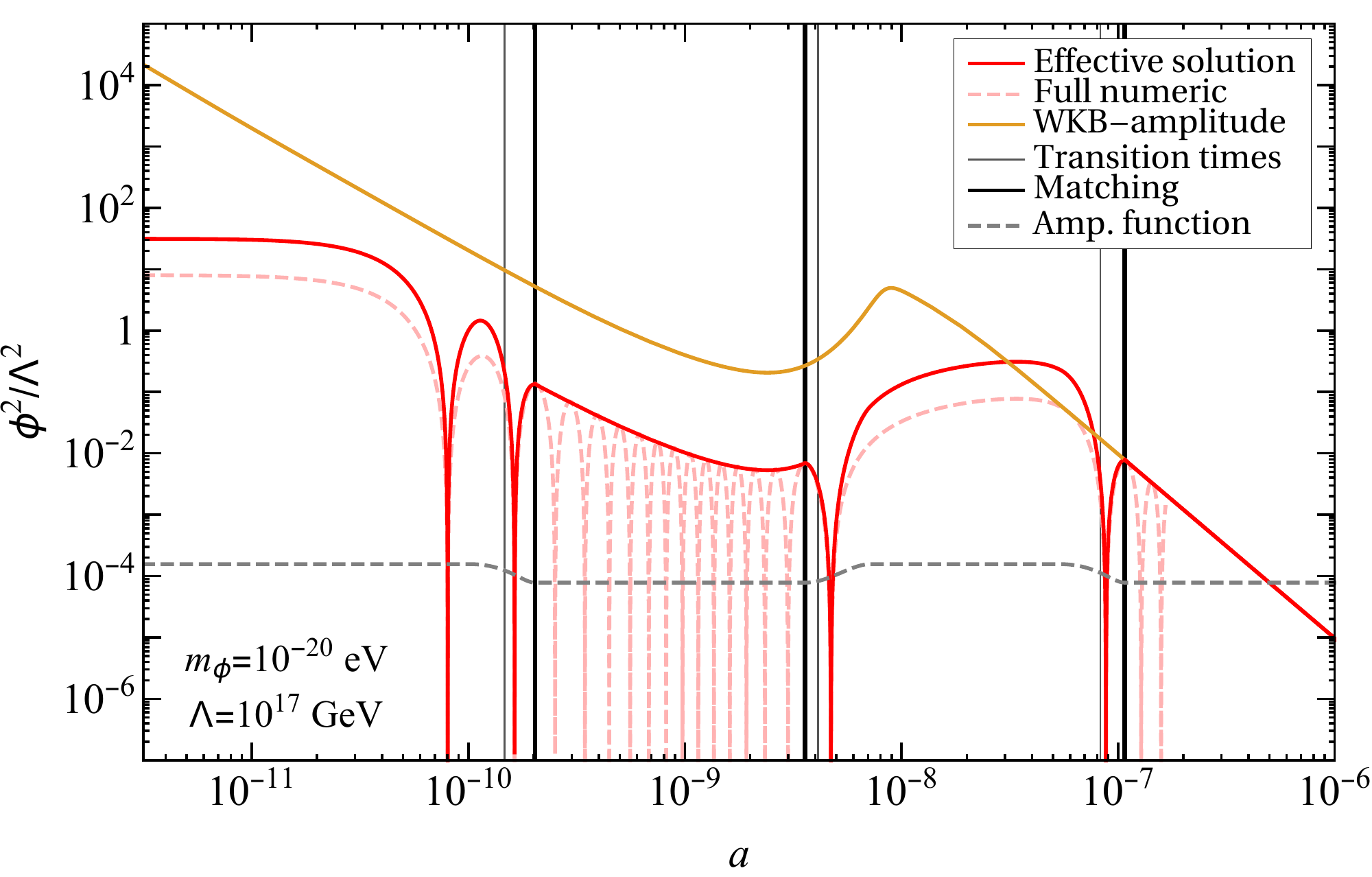}
\caption{Same as fig.~\ref{fig:DM density} (right), but with transition times marked with vertical lines and the amplification function visualized by the red dashed line. The amplification function (gray, dashed), which takes values in the range between 1 and 2, is shown here with an artificial amplitude for visualization. The effective solution (red, solid) is the patching between the full numeric solution (red, dashed) and the WKB-amplitude (orange, solid) for the rapidly-oscillating regions.}
\label{fig:evolutionPlotDetailed}
\end{figure}

In this appendix we describe our patching procedure for the cosmological evolution of the DM. The patching procedure is required because of rapid oscillations, which are present both in the bare mass regime (B) and in the induced mass regime (I), when positive couplings are considered.
These oscillations can make calculating full numerical solutions computationally expensive. 
Alternatively, the WKB-approximation allows us to accurately reproduce the evolution in the regime where the field is rapidly oscillating. Since the WKB-approximation is only valid in the rapidly oscillating regime, we must patch between the full numerical solution and the WKB-approximation. We implement the patching as follows:
\begin{enumerate}
\item Compare the total effective mass $m_{\phi,{\rm eff}}$ to the Hubble parameter to determine if and when the Hubble friction becomes relevant. Specifically, we require that the Hubble parameter $H$ is bigger than $0.2\ m_{\phi,{\rm eff}}$ before the numerical solution is applied. This determines when to transition between the full numerical solution and the WKB-approximation. Note that in the case of intermediate fast regimes, triggered by the induced mass, there will be several transitions.\label{transition time step}
\item Numerically solve the equations of motion from some initial time ($ \log(a)=-12 $) and until the latest transition time (i.e. the highest value of $a$) found above.
\item In order to ensure the correct normalization, we need to match the solutions at the peaks of the oscillations. We therefore numerically search for peaks in the oscillations near the desired transition times. We perform the searches away from the slowly varying regions to ensure that we always find peaks in the fast regime.
\item The amplitude of the WKB solution is fixed to the present day DM density. We can therefore fix the amplitude of the numerical solution by matching the amplitude of the numerical solution to the WKB solution at the peaks found above. If an intermediate fast regime is present, then the process is reversed and the amplitude of the intermediate WKB solution is matched to a peak of the numerics. 
\item For the BBN analysis we need the value of $\phi^2/\Lambda^2$ averaged over fast oscillations. This is smaller by a factor 2 than the actual envelope of the field. As we normalize our WKB amplitude to track the {\em average} of the numerical solution, we have to multiply the numerical solution by a factor 2 when transitioning from the WKB to the Hubble friction dominated regime.
To prevent the evolution from being discontinuous we gradually impose the factor of 2 by turning it on or off over a period of a Hubble time.
\end{enumerate}
We illustrate the matching procedure in fig.~\ref{fig:evolutionPlotDetailed}, indicating the various types of solutions: full numeric, WKB-amplitude, and the effective solution. Here, we show the different transition times, indicated by thin black vertical lines, which are found by comparing the total mass and the Hubble parameter. The solutions are matched at the peaks, marked with thick, black vertical lines.  The amplification function, which takes the factor of 2 from averaging into account, is also visualized with a gray dotted line, although the function is shown with an artificial amplitude for visualization purposes. The effective solution used for the BBN analysis is shown as a red, solid curve. It results from the patching of the full numeric solution  and the WKB-amplitude for the rapidly-oscillating regions.

For negative coupling the patching procedure is similar, except that in the induced mass regime (I) the solution is exponentially growing, instead of oscillating. In this case we do not use the WKB approximation in the I regime, but solve the equation exactly.

\section{Expansion of the universe in Einstein and Jordan frames}
\label{app:Einstein-Jordan map}

Here we summarize the equations of motion in the Einstein and Jordan
frames, discuss the relations between various quantities in the two
frames and derive how the presence of ULDM perturbs the expansion of the universe during BBN.
We keep only the leading order corrections in $\alpha(\phi)$. The
general expressions can be found in Ref.~\cite{Coc:2006rt}.

From the Einstein frame Lagrangian (\ref{EinsteinLag}) we obtain the
Friedmann equation,
\begin{subequations}
\begin{equation}
\label{EinsFried}
3M_{\rm pl}^2 H^2=\rho_{\rm
  SM}+\frac{1}{2}\bigg(\frac{d\phi}{dt}\bigg)^2
+\frac{m_\phi^2\phi^2}{2}\;,\qquad \text{where}~H\equiv\frac{d\ln a}{dt}\;,
\end{equation}
the scalar field equation,
\begin{equation}
\frac{d^2\phi}{dt^2}+3H\frac{d\phi}{dt}+m_\phi^2\phi
+\frac{d\alpha}{d\phi}\Theta_{\rm SM} =0\;,
\end{equation}
and the SM energy conservation,
\begin{equation}
\label{EinstCons}
\frac{d\rho_{\rm SM}}{dt}+3H(\rho_{\rm SM}+p_{\rm SM})
=\frac{d\alpha}{dt} \Theta_{\rm SM}\;.
\end{equation}
\end{subequations}
This system must be complemented by the equation of state relating
$p_{\rm SM}$ to $\rho_{\rm SM}$ and $\phi$, where the dependence on $\phi$
appears due to the $\phi$-modulation of the SM parameters
(see below). 

The corresponding equations in the Jordan frame follow from the
Lagrangian (\ref{JordanLag1}),
\begin{subequations}
\begin{align}
&3M_{\rm pl}^2(1-2\alpha)\bar H^2=\bar\rho_{\rm SM}
+6M_{\rm pl}^2\bar H\frac{d\alpha}{d\bar t}-
3M_{\rm pl}^2\bigg(\frac{d\alpha}{d\bar t}\bigg)^2\notag\\
&\qquad\qquad\qquad+\frac{1-2\alpha}{2}\bigg(\frac{d\phi}{d\bar t}\bigg)^2
+\frac{(1-4\alpha)m_\phi^2\phi^2}{2}\;,\qquad\text{where}~
\bar H\equiv\frac{d\ln \bar a}{d\bar t}\;,
\label{JordFried}\\
&\bigg[1-2\alpha-6M_{\rm pl}^2\bigg(\frac{d\alpha}{d\phi}\bigg)^2\bigg]
\bigg(\frac{d^2\phi}{d\bar t^2}+3\bar H\frac{d\phi}{d\bar t}\bigg)
+6M_{\rm pl}^2\frac{d\alpha}{d\phi}\bigg(\frac{d\bar H}{d\bar t}+2\bar
H^2\bigg)\notag\\
&\qquad\qquad\qquad-\bigg(\frac{d\alpha}{d\phi}+6M_{\rm pl}^2\frac{d\alpha}{d\phi}
\frac{d^2\alpha}{d\phi^2}\bigg)\bigg(\frac{d\phi}{d\bar t}\bigg)^2
+\bigg(1-4\alpha-2\phi\frac{d\alpha}{d\phi}\bigg) m_\phi^2\phi=0\;,\\
&\frac{d\bar\rho_{\rm SM}}{d\bar t}+3\bar H(\bar\rho_{\rm SM}+\bar
p_{\rm SM})=0\;,
\label{JordCons}
\end{align}
\end{subequations}
where we have marked the Jordan quantities with a bar to differentiate
them from their Einstein counterparts. Notice that the SM energy is
separately conserved in the Jordan frame because in this frame there is no direct
coupling between SM and $\phi$. For the same reason, the SM pressure
obeys the standard $\phi$-independent equation of state, 
$\bar p_{\rm SM}=p^{(0)}(\bar \rho_{\rm SM})$. This implies that $\bar
\rho_{\rm SM}$ is the same function of the Jordan scale factor $\bar
a$ (but not the time~$\bar t$~!) as in the standard cosmology, 
\begin{equation}
\label{JordStand}
\bar \rho_{\rm SM}(\bar a)=\rho^{(0)}_{SM}(\bar a)\;.
\end{equation}

To establish the map between the Einstein and Jordan frames,
we use the basic relation~(\ref{effmetr}). This implies the connection
between the proper times and scale factors,
\begin{gather}
{\rm d}\bar t = (1+\alpha) {\rm d}t,\qquad\qquad
\bar a = (1+\alpha) a\;.
\label{aEtoaJ}
\end{gather}
These, in turn, give the connection between the Hubble parameters,
\begin{gather}
\bar H = H(1 - \alpha)+\frac{d\alpha}{d t}
=H\bigg(1-\alpha+\frac{d \alpha}{d\ln a}\bigg)\;.
\label{HJ_first}
\end{gather}
Note that in this expression $\bar H$ on the l.h.s. is taken at the
Jordan time $\bar t$ (or the scale factor $\bar a$), whereas the r.h.s
is evaluated at the corresponding Einstein time $t$ (scale factor~$a$).

Next we consider the SM energy density and pressure. Recall that these
are defined as,
\begin{equation}
\rho_{\rm SM}=T_{{\rm SM}\,\mu\nu}u^\mu u^\nu\;,~~~~~~~
p_{\rm SM}=\frac{1}{3}T_{{\rm SM}\,\mu\nu}(u^\mu u^\nu-g^{\mu\nu})\;,
\end{equation}
where 
\begin{equation}
T_{{\rm SM}\,\mu\nu}=\frac{2}{\sqrt{-g}}\frac{\delta S_{\rm
    SM}}{\delta g^{\mu\nu}}
\end{equation}
is the SM energy-momentum tensor and $u^\mu$ is a time-directed vector
with unit norm, $u^\mu u^\nu g_{\mu\nu}=1$. Similar expressions hold
in the Jordan frame with the replacement $g_{\mu\nu}\mapsto \bar
g_{\mu\nu}$. Using again (eq.~\ref{effmetr}) we obtain,
\begin{gather}
\bar T_{{\rm SM}\,\mu\nu}=(1-2\alpha)  T_{{\rm
    SM}\,\mu\nu}~,\qquad\qquad
\bar u^\mu =(1-\alpha)u^\mu\;,\\
\bar\rho_{\rm SM}=(1-4\alpha)\rho_{\rm SM}~,\qquad\qquad
\bar p_{\rm SM}=(1-4\alpha) p_{\rm SM}\;.
\label{rhoPEJ}
\end{gather}
The last expressions imply the Einstein frame equation of state,
\begin{equation}
\label{EinsEOS}
p_{\rm SM}=\bigg(1+4\alpha(\phi)-4\alpha(\phi)\frac{d\ln
  p^{(0)}(\rho_{\rm SM})}{d\ln \rho_{\rm SM}}\bigg) p^{(0)}(\rho_{\rm SM})\;.
\end{equation}
Notice that this expression is in general $\phi$-dependent. Only for a
linear equation of state $p^{(0)}(\rho_{\rm SM})=w \rho_{\rm SM}$,
with constant $w$, the $\phi$-dependence drops out and one recovers
the standard relation $p_{\rm SM}=w\rho_{\rm SM}$.

Let us now discuss how the ULDM perturbs the expansion of the universe
in the two frames during BBN. In the Einstein frame the Hubble rate obeys the
Fiedmann eq.~(\ref{EinsFried})
where the last two terms represent the DM energy density. We argue that the latter is negligible. The DM density is comparable
to that of radiation at the scale factor 
$a_{\rm eq}\sim 10^{-4}$. If the DM dynamics
is dominated by the bare mass, the ratio $\rho_\phi/\rho_{\rm SM}$
scales like $a/a_{\rm eq}$ and is less than $10^{-4}$ at the BBN
epoch, which is too small to affect BBN in an observable way.
In the Hubble friction regime, when the value of $\phi$ is
frozen, the ratio $\rho_\phi/\rho_{\rm SM}$ is suppressed even
further. It remains to check what happens when the induced mass
dominates. In this case we can use the WKB solutions
(\ref{expgrowth}), (\ref{WKB}) to obtain
\begin{equation}
\label{rhophiest}
\frac{\rho_\phi}{\rho_{\rm SM}}\simeq
\frac{|m_{\phi,\rm{eff}}^2|\phi^2}{2\rho_{\rm SM}}
=\frac{\Theta_{\rm SM}\phi^2}{\sqrt{2}\Lambda^2\rho_{\rm SM}}
\simeq \frac{1}{\sqrt{2}}|\alpha(\phi)|\Sigma\;.
\end{equation} 
This expression is doubly suppressed by the small quantities
$|\alpha(\phi)|$ and the small ratio $\Sigma$ introduced in
eq.~(\ref{eq:Sigma}). We saw in the main text that the leading 
effect on BBN is
of order $O\big(\alpha(\phi)\big)$, so the contributions of the form
\eqref{rhophiest} can be neglected. 

Now we need to determine the effect of the ULDM coupling on the SM
energy density $\rho_{\rm SM}$. This is non-trivial, since
the SM energy is not conserved due to the direct
DM coupling, see
eq.~(\ref{EinstCons}). Moreover, the equation of state relating the 
SM pressure to
the energy density is modified, eq.~(\ref{EinsEOS}). To overcome these
complications, we use the map to the Jordan frame, where the energy
density has the standard dependence on the scale factor
(\ref{JordStand}).  
Using the first of eqs.~(\ref{rhoPEJ}), we obtain,
\begin{align}
\rho_{\rm SM}(a)&=(1+4\alpha) \rho_{\rm SM}^{(0)}\big((1+\alpha) a\big)\notag\\
&=\bigg(1+4\alpha+\alpha\frac{d\ln \rho_{\rm SM}^{(0)}(a)}{d\ln
  a}\bigg)\rho_{\rm SM}^{(0)}(a)\;,
\label{rhoSM1}
\end{align}
where we have expanded to the linear order in $\alpha$.
To simplify this expression we use the energy conservation of the
standard cosmology which can be written as follows,
\begin{equation}
\label{standCons}
\frac{d\ln \rho_{\rm SM}^{(0)}(a)}{d\ln a}+4-\Sigma=0\;.
\end{equation}
Substitution into eq.~(\ref{rhoSM1}) yields,
\begin{equation}
\rho_{\rm SM}(a)=(1+\alpha\Sigma) \rho_{\rm SM}^{(0)}(a)\;.
\label{rhoSM2}
\end{equation}
We see that, though $\rho_{\rm SM}$ changes with respect to the
standard cosmology, the deviation is doubly suppressed by $\alpha$ and
$\Sigma$. Thus, we can omit it within our approximation. Substituting
into the Fiedmann equation, we conclude that, up to terms of order
$O(\alpha\Sigma)$, the Hubble rate in the Einstein frame is described
by the same function of the scale factor, as in the standard
cosmology. In this way we arrive to eqs.~(\ref{rhoHEins}) from the main text.

The Hubble rate in the Jordan frame is determined by
eq.~(\ref{JordFried}). Using the same arguments as for the
$\phi$-density in the Einstein frame, one can show that the two terms
in the second line of this equation are negligible. Recalling also
that the dependence of the SM density on the Jordan scale factor is
standard, we immediately obtain,
\begin{equation}
\label{JordH1}
\bar H(\bar a)=H^{(0)}(\bar a)(1+\alpha)+\frac{d\alpha}{d\bar t}\;,
\end{equation}
which is equivalent to eq.~(\ref{JordH}) from the main text. Notice
that this expression is consistent with eq.~(\ref{HJ_first}) upon
taking into account the relations,
\[
H(a)\simeq H^{(0)}(a)=H^{(0)}(\bar a)\bigg(1-\alpha \frac{d\ln
  H^{(0)}}{d\ln a}\bigg)\simeq H^{(0)}(\bar a)(1+2\alpha)\;.
\]

\end{appendices}

 \bibliographystyle{JHEP}
 \bibliography{BBN.bib}
\end{document}